\documentclass[%
reprint,
superscriptaddress,
 amsmath,amssymb,
 aps,
 prc,
 showkeys,
 preprintnumbers,
floatfix,
]{revtex4-2}

\usepackage{graphicx}
\usepackage{dcolumn}
\newcolumntype{d}[1]{D{.}{.}{-1}} 
\usepackage{latexsym}
\usepackage{amsmath} 
\usepackage{url}
\usepackage{natbib}
\usepackage[running]{lineno}
\usepackage{color}
\usepackage{comment}
\usepackage{mathtools}
\usepackage{framed}
\usepackage{diagbox}
\usepackage{multirow}
\usepackage{hyperref}
\usepackage[margin=20mm]{geometry}
\usepackage{siunitx}
\usepackage{subfigure}
\usepackage{makecell}

\definecolor{forest}{RGB}{0, 153, 51}

\newcolumntype{C}[1]{>{\centering\arraybackslash}m{#1}}

\usepackage[normalem]{ulem}
 
\begin{document}

\preprint{Phys. Rev. C}

\title{Measurements of $\gamma_v p \to \pi^+ \pi^- p'$ Cross Sections with the CLAS Detector for $Q^{2}$ from 2.0--5.0~GeV$^{2}$ and $W$ from 1.400--2.125~GeV}



\newcommand*{\ANL}{Argonne National Laboratory, Argonne, Illinois 60439}
\newcommand*{\ANLindex}{1}
\affiliation{\ANL}
\newcommand*{\ASU}{Arizona State University, Tempe, Arizona 85287-1504}
\newcommand*{\ASUindex}{2}
\affiliation{\ASU}
\newcommand*{\BNL}{Brookhaven National Laboratory, Upton, New York 11973-5000}
\newcommand*{\BNLindex}{3}
\affiliation{\BNL}
\newcommand*{\CMU}{Carnegie Mellon University, Pittsburgh, Pennsylvania 15213}
\newcommand*{\CMUindex}{4}
\affiliation{\CMU}
\newcommand*{\CUA}{Catholic University of America, Washington, D.C. 20064}
\newcommand*{\CUAindex}{5}
\affiliation{\CUA}
\newcommand*{\SACLAY}{IRFU, CEA, Universit\'e Paris-Saclay, F-91191 Gif-sur-Yvette, France}
\newcommand*{\SACLAYindex}{6}
\affiliation{\SACLAY}
\newcommand*{\CNU}{Christopher Newport University, Newport News, Virginia 23606}
\newcommand*{\CNUindex}{7}
\affiliation{\CNU}
\newcommand*{\UCONN}{University of Connecticut, Storrs, Connecticut 06269}
\newcommand*{\UCONNindex}{8}
\affiliation{\UCONN}
\newcommand*{\DUKE}{Duke University, Durham, North Carolina 27708-0305}
\newcommand*{\DUKEindex}{9}
\affiliation{\DUKE}
\newcommand*{\DUQUESNE}{Duquesne University, 600 Forbes Avenue, Pittsburgh, Pennsylvania 15282}
\newcommand*{\DUQUESNEindex}{10}
\affiliation{\DUQUESNE}
\newcommand*{\FU}{Fairfield University, Fairfield, Connecticut 06824}
\newcommand*{\FUindex}{11}
\affiliation{\FU}
\newcommand*{\FERRARAU}{Universita' di Ferrara, 44121 Ferrara, Italy}
\newcommand*{\FERRARAUindex}{12}
\affiliation{\FERRARAU}
\newcommand*{\FIU}{Florida International University, Miami, Florida 33199}
\newcommand*{\FIUindex}{13}
\affiliation{\FIU}
\newcommand*{\FSU}{Florida State University, Tallahassee, Florida 32306}
\newcommand*{\FSUindex}{14}
\affiliation{\FSU}
\newcommand*{\GWUI}{The George Washington University, Washington, DC 20052}
\newcommand*{\GWUIindex}{15}
\affiliation{\GWUI}
\newcommand*{\GSIFFN}{GSI Helmholtzzentrum fur Schwerionenforschung GmbH, D-64291 Darmstadt, Germany}
\newcommand*{\GSIFFNindex}{16}
\affiliation{\GSIFFN}
\newcommand*{\ORSAY}{Universit\'e Paris-Saclay, CNRS/IN2P3, IJCLab, 91405 Orsay, France}
\newcommand*{\ORSAYindex}{17}
\affiliation{\ORSAY}
\newcommand*{\INFNCAT}{INFN, Sezione di Catania, 95123 Catania, Italy}
\newcommand*{\INFNCATindex}{18}
\affiliation{\INFNCAT}
\newcommand*{\INFNFE}{INFN, Sezione di Ferrara, 44100 Ferrara, Italy}
\newcommand*{\INFNFEindex}{19}
\affiliation{\INFNFE}
\newcommand*{\INFNFR}{INFN, Laboratori Nazionali di Frascati, 00044 Frascati, Italy}
\newcommand*{\INFNFRindex}{20}
\affiliation{\INFNFR}
\newcommand*{\INFNGE}{INFN, Sezione di Genova, 16146 Genova, Italy}
\newcommand*{\INFNGEindex}{21}
\affiliation{\INFNGE}
\newcommand*{\INFNRO}{INFN, Sezione di Roma Tor Vergata, 00133 Rome, Italy}
\newcommand*{\INFNROindex}{22}
\affiliation{\INFNRO}
\newcommand*{\INFNTUR}{INFN, Sezione di Torino, 10125 Torino, Italy}
\newcommand*{\INFNTURindex}{23}
\affiliation{\INFNTUR}
\newcommand*{\INFNPAV}{INFN, Sezione di Pavia, 27100 Pavia, Italy}
\newcommand*{\INFNPAVindex}{24}
\affiliation{\INFNPAV}
\newcommand*{\JMU}{James Madison University, Harrisonburg, Virginia 22807}
\newcommand*{\JMUindex}{25}
\affiliation{\JMU}
\newcommand*{\MIT}{Massachusetts Institute of Technology, Cambridge, Massachusetts 02139-4307}
\newcommand*{\MITindex}{26}
\affiliation{\MIT}
\newcommand*{\MISS}{Mississippi State University, Mississippi State, Mississippi 39762-5167}
\newcommand*{\MISSindex}{27}
\affiliation{\MISS}
\newcommand*{\UNH}{University of New Hampshire, Durham, New Hampshire 03824-3568}
\newcommand*{\UNHindex}{28}
\affiliation{\UNH}
\newcommand*{\NMSU}{New Mexico State University, PO Box 30001, Las Cruces, New Mexico 88003}
\newcommand*{\NMSUindex}{29}
\affiliation{\NMSU}
\newcommand*{\NSU}{Norfolk State University, Norfolk, Virginia 23504}
\newcommand*{\NSUindex}{30}
\affiliation{\NSU}
\newcommand*{\OHIOU}{Ohio University, Athens, Ohio 45701}
\newcommand*{\OHIOUindex}{31}
\affiliation{\OHIOU}
\newcommand*{\ODU}{Old Dominion University, Norfolk, Virginia 23529}
\newcommand*{\ODUindex}{32}
\affiliation{\ODU}
\newcommand*{\JLUGiessen}{II Physikalisches Institut der Universitaet Giessen, 35392 Giessen, Germany}
\newcommand*{\JLUGiessenindex}{33}
\affiliation{\JLUGiessen}
\newcommand*{\RPI}{Rensselaer Polytechnic Institute, Troy, New York 12180-3590}
\newcommand*{\RPIindex}{34}
\affiliation{\RPI}
\newcommand*{\ROMAII}{Universita' di Roma Tor Vergata, 00133 Rome Italy}
\newcommand*{\ROMAIIindex}{35}
\affiliation{\ROMAII}
\newcommand*{\SDU}{Shandong University, Qingdao, Shandong 266237, China}
\newcommand*{\SDUindex}{36}
\affiliation{\SDU}
\newcommand*{\MSU}{Skobeltsyn Institute of Nuclear Physics, Lomonosov Moscow State University, 119234 Moscow, Russia}
\newcommand*{\MSUindex}{37}
\affiliation{\MSU}
\newcommand*{\SCAROLINA}{University of South Carolina, Columbia, South Carolina 29208}
\newcommand*{\SCAROLINAindex}{38}
\affiliation{\SCAROLINA}
\newcommand*{\TEMPLE}{Temple University,  Philadelphia, Pennsylvania 19122}
\newcommand*{\TEMPLEindex}{39}
\affiliation{\TEMPLE}
\newcommand*{\JLAB}{Thomas Jefferson National Accelerator Facility, Newport News, Virginia 23606}
\newcommand*{\JLABindex}{40}
\affiliation{\JLAB}
\newcommand*{\ULS}{Universidad de La Serena, Avda. Juan Cisternas 1200, La Serena, Chile}
\newcommand*{\ULSindex}{41}
\affiliation{\ULS}
\newcommand*{\UTFSM}{Universidad T\'{e}cnica Federico Santa Mar\'{i}a, Casilla 110-V Valpara\'{i}so, Chile}
\newcommand*{\UTFSMindex}{42}
\affiliation{\UTFSM}
\newcommand*{\BRESCIA}{Universit`{a} degli Studi di Brescia, 25123 Brescia, Italy}
\newcommand*{\BRESCIAindex}{43}
\affiliation{\BRESCIA}
\newcommand*{\GLASGOW}{University of Glasgow, Glasgow G12 8QQ, United Kingdom}
\newcommand*{\GLASGOWindex}{44}
\affiliation{\GLASGOW}
\newcommand*{\YORK}{University of York, York YO10 5DD, United Kingdom}
\newcommand*{\YORKindex}{45}
\affiliation{\YORK}
\newcommand*{\VIRGINIA}{University of Virginia, Charlottesville, Virginia 22901}
\newcommand*{\VIRGINIAindex}{46}
\affiliation{\VIRGINIA}
\newcommand*{\WM}{College of William and Mary, Williamsburg, Virginia 23187-8795}
\newcommand*{\WMindex}{47}
\affiliation{\WM}
\newcommand*{\YEREVAN}{Yerevan Physics Institute, 375036 Yerevan, Armenia}
\newcommand*{\YEREVANindex}{48}
\affiliation{\YEREVAN}
 
\newcommand*{\NOWINFNRO}{INFN, Sezione di Roma Tor Vergata, 00133 Rome, Italy}
\newcommand*{\NOWISU}{Idaho State University, Pocatello, Idaho 83209}
\newcommand*{\NOWUCONN}{University of Connecticut, Storrs, Connecticut 06269}

\author {A. Trivedi} 
\affiliation{\SCAROLINA}
\author {R.W.~Gothe} 
\affiliation{\SCAROLINA}
\author {E. Phelps} 
\affiliation{\SCAROLINA}
\author {V.I. Mokeev} 
\affiliation{\SCAROLINA}
\affiliation{\JLAB}
\author {D.S.~Carman} 
\affiliation{\JLAB}


\author {P.~Achenbach} 
\affiliation{\CNU}
\author {J. S. Alvarado} 
\affiliation{\ORSAY}
\author {M.J.~Amaryan}
\affiliation{\ODU}
\author {N.A.~Baltzell} 
\affiliation{\JLAB}
\author {L. Barion} 
\affiliation{\INFNFE}
% ...
\collaboration{The CLAS Collaboration}
\noaffiliation

\date{\today}

\begin{abstract}
Observables in the electroproduction of the $\pi^+\pi^-p$ reaction channel off the proton that are sensitive to the polarization of the virtual photon are presented for the first time in addition to the extraction of the nine single-differential and fully integrated cross sections within the kinematics area of 2.0~GeV$^2 < Q^2 < 5.0$~GeV$^2$ and 1.400~GeV $< W < 2.125$~GeV measured with the CLAS detector in Hall B at Jefferson Lab. The extraction of the unpolarized cross sections has been considerably improved in comparison with previously published results from this same dataset, offering finer binning over the five-dimensional hadronic reaction phase space, and including essential advances in the acceptance evaluation by implementing a new technique to stabilize the cross section extraction and minimize the systematic uncertainty associated with the simulation statistics. These improvements are of particular importance for the extraction of the $\gamma_v p N^*$ electrocouplings from these data.
\end{abstract}

\maketitle

\newcommand{\fullrctn}{$ep \to e'\pi^+\pi^-p'$}
\newcommand{\vprctn}{$\gamma_vp \to \pi^+\pi^-p'$ }
\newcommand{\fullrctnthreepion}{$ep \to e'\pi^+\pi^-\pi^0p'$}
\newcommand{\threepiPSsim}{$3\pi$ simulation}
\newcommand{\twopisim}{$2\pi$ simulation}
\newcommand{\threepi}{$3\pi$}
\newcommand{\twopi}{$2\pi$}

\newcommand{\defeq}{$\vcentcolon=$}
\newcommand{\ignore}[1]{}
\newcommand{\deltatcut}{$\Delta t$-cut}
\newcommand{\qw}{$Q^2-W$}
\newcommand{\qwbin}{$Q^2-W$ bin}
\newcommand{\qwbins}{$Q^2-W$ bins}

\newcommand{\GeVsq}{GeV\textsuperscript{2}}

\newcommand{\RAwOO}{${R2_{T}}^{00} + {R2_{L}}^{00}$}
\newcommand{\RBwOO}{${R2_{LT}}^{c,00}$}
\newcommand{\RCwOO}{${R2_{TT}}^{c,00}$}
\newcommand{\RDwOO}{${R2_{LT}}^{s,00}$}
\newcommand{\REwOO}{${R2_{TT}}^{s,00}$}

\newcommand{\RA}{${R2_{T}}^{X_{ij}}_{\phi_{i}} + {R2_{L}}^{X_{ij}}_{\phi_{i}}$}
\newcommand{\RB}{${R2_{LT}}^{c,X_{ij}}_{\phi_{i}}$}
\newcommand{\RC}{${R2_{TT}}^{c,X_{ij}}_{\phi_{i}}$}
\newcommand{\RD}{${R2_{LT}}^{s,X_{ij}}_{\phi_{i}}$}
\newcommand{\RE}{${R2_{TT}}^{s,X_{ij}}_{\phi_{i}}$}

\newcommand{\RAapndx}{${R2_{T}}^{\alpha_{[p_{f}\pi^{+}][p\pi^{-}]}}_{\phi_{\pi^{-}}} + {R2_{L}}^{\alpha_{[p_{f}\pi^{+}][p\pi^{-}]}}_{\phi_{\pi^{-}}}$}
\newcommand{\RBapndx}{${R2_{LT}}^{c,\alpha_{[p_{f}\pi^{+}][p\pi^{-}]}}_{\phi_{\pi^{-}}}$}
\newcommand{\RCapndx}{${R2_{TT}}^{c,\alpha_{[p_{f}\pi^{+}][p\pi^{-}]}}_{\phi_{\pi^{-}}}$}
\newcommand{\RDapndx}{${R2_{LT}}^{s,\alpha_{[p_{f}\pi^{+}][p\pi^{-}]}}_{\phi_{\pi^{-}}}$}
\newcommand{\REapndx}{${R2_{TT}}^{s2,\alpha_{[p_{f}\pi^{+}][p\pi^{-}]}}_{\phi_{\pi^{-}}}$}

\newcommand{\illusqwbin}{$Q^{2}$=[2.40 \GeVsq, \allowbreak3.00 \GeVsq) and $W$=[1.725 GeV, \allowbreak1.750 GeV)}

\newcommand{\frmla}{
\begin{align} \label{eqn2Dphiproj}
\left(\frac{d^2\sigma}{ {dX_{ij}d\phi_{i}} }\right)=
\begin{split}
&{R2_{T}}^{X_{ij}}_{\phi_{i}} + {R2_{L}}^{X_{ij}}_{\phi_{i}} + \\ 
&{R2_{LT}}^{c,X_{ij}}_{\phi_{i}}\cos\phi_{i} +  {R2_{TT}}^{c,X_{ij}}_{\phi_{i}}\cos2\phi_{i} + \\
&\delta_{X_{ij}\alpha_{i}}\left({R2_{LT}}^{s,\alpha_{i}}_{\phi_{i}}\sin\phi_{i} + {R2_{TT}}^{s,\alpha_{i}}_{\phi_{i}}\sin2\phi_{i}\right)
\end{split}
\end{align}
}

\newcommand{\frmlaintg}{
\begin{equation} \label{eqn2Dphiprojintg}
\begin{split}
\frac{d\sigma}{ {dX_{ij}} }  = \left({R2_{T}}^{X_{ij}}_{\phi_{i}} + {R2_{L}}^{X_{ij}}_{\phi_{i}}\right)\cdot2\pi
\end{split}
\end{equation}
}

\newcommand{\frmlayield}{
\begin{equation} \label{eqn2Dphiprojyield}
\begin{split}
\Delta^2N(\Delta X_{ij},\Delta\phi_{i})  = \int\displaylimits_{\Delta X_{ij}\Delta\phi_{i}} \Big({R2_{T}}^{X_{ij}}_{\phi_{i}} + {R2_{L}}^{X_{ij}}_{\phi_{i}} + {R2_{LT}}^{c,X_{ij}}_{\phi_{i}}\cos\phi_{i} + \\ {R2_{TT}}^{c,X_{ij}}_{\phi_{i}}\cos2\phi_{i} + \delta_{X_{ij}\alpha_{i}}\left({R2_{LT}}^{s,\alpha_{i}}_{\phi_{i}}\sin\phi_{i} + {R2_{TT}}^{s,\alpha_{i}}_{\phi_{i}}\sin2\phi_{i}\right) \Big) dX_{ij}d\phi_{i}\cdot NF
\end{split}
\end{equation}
}

\newcommand{\polobsnf}{
\begin{equation} \label{eqnpolobsnf}
NF = L\cdot\Gamma_{v}\cdot\Delta Q^{2}\cdot\Delta W\cdot R
\end{equation}
}

\newcommand{\fitfnctn}{
\begin{equation} \label{eqnfitfnctn}
\begin{split}
f = A + B\cos\phi + C\cos2\phi + \delta_{X_{ij}\alpha_{i}}\left(D\sin\phi + E\sin2\phi\right)
\end{split}
\end{equation}
}

\newcommand{\fitCF}{
\begin{align} \label{eqnfitCF}
\begin{split}
CF_{A} &= \frac{1}{\int\displaylimits_{\Delta\phi_{i}} d\phi_{i}} \\
CF_{B} &= \frac{\cos\phi_{i}^{\text{binc}}}{\int\displaylimits_{\Delta\phi_{i}} \cos\phi_{i}d\phi_{i}} \\
CF_{C} &= \frac{\cos2\phi_{i}^{\text{binc}}}{\int\displaylimits_{\Delta\phi_{i}} \cos2\phi_{i}d\phi_{i}} \\
CF_{D} &= \frac{\sin\phi_{i}^{\text{binc}}}{\int\displaylimits_{\Delta\phi_{i}} \sin\phi_{i}d\phi_{i}} \\
CF_{E} &= \frac{\sin2\phi_{i}^{\text{binc}}}{\int\displaylimits_{\Delta\phi_{i}} \sin2\phi_{i}d\phi_{i}}
\end{split}
\end{align}
}

\newcommand{\polobsfinalextrct}{
\begin{align} \label{eqnpolobsfinalextrct}
\begin{split}
{R2_{T}}^{X_{ij}}_{\phi_{i}} + {R2_{L}}^{X_{ij}}_{\phi_{i}} &= {\frac{A\cdot CF_{A}}{L\cdot\Gamma_{v}\cdot\Delta Q^{2}\cdot\Delta W\cdot\Delta X_{ij}\cdot R}}\cdot \mathcal{E} \\
{R2_{LT}}^{c,X_{ij}}_{\phi_{i}} &= {\frac{B\cdot CF_{B}}{L\cdot\Gamma_{v}\cdot\Delta Q^{2}\cdot\Delta W\cdot\Delta X_{ij}\cdot R}}\cdot \mathcal{E} \\
{R2_{TT}}^{c,X_{ij}}_{\phi_{i}} &= {\frac{C\cdot CF_{C}}{L\cdot\Gamma_{v}\cdot\Delta Q^{2}\cdot\Delta W\cdot\Delta X_{ij}\cdot R}}\cdot \mathcal{E} \\
{R2_{LT}}^{s,X_{ij}}_{\phi_{i}} &= {\frac{D\cdot CF_{D}}{L\cdot\Gamma_{v}\cdot\Delta Q^{2}\cdot\Delta W\cdot\Delta X_{ij}\cdot R}}\cdot \mathcal{E} \\
{R2_{TT}}^{s2,X_{ij}}_{\phi_{i}}&= {\frac{E\cdot CF_{E}}{L\cdot\Gamma_{v}\cdot\Delta Q^{2}\cdot\Delta W\cdot\Delta X_{ij}\cdot R}} \cdot \mathcal{E}
\end{split}
\end{align}
}

\newcommand{\polobsfinalextrctexmplalpha}{
\begin{align} \label{polobsfinalextrctexmplalpha}
\begin{split}
{R2_{T}}^{\alpha_{[p_{f}\pi^{+}][p\pi^{-}]}}_{\phi_{\pi^{-}}} + {R2_{L}}^{\alpha_{[p_{f}\pi^{+}][p\pi^{-}]}}_{\phi_{\pi^{-}}} &= {\frac{2622.22\cdot 1.59155}{\num{28.18e9}\cdot0.000117\cdot0.6\cdot0.025\cdot0.628318}}\cdot1.0183\\ 
=0.158949\;\frac{\mu b}{\text{rad}}\\
{R2_{LT}}^{c,\alpha_{[p_{f}\pi^{+}][p\pi^{-}]}}_{\phi_{\pi^{-}}} &= {\frac{-657.15\cdot 1.61803}{\num{28.18e9}\cdot0.000117\cdot0.6\cdot0.025\cdot0.628318}}\cdot1.0183 \\
=-0.040496\;\frac{\mu b}{\text{rad}}\\
{R2_{TT}}^{c,\alpha_{[p_{f}\pi^{+}][p\pi^{-}]}}_{\phi_{\pi^{-}}} &= {\frac{54.45\cdot 1.70131}{\num{28.18e9}\cdot0.000117\cdot0.6\cdot0.025\cdot0.628318}}\cdot1.0183\\
0.003528\;\frac{\mu b}{\text{rad}}\\
{R2_{LT}}^{s,\alpha_{[p_{f}\pi^{+}][p\pi^{-}]}}_{\phi_{\pi^{-}}} &= {\frac{265.50\cdot 1.61803}{\num{28.18e9}\cdot0.000117\cdot0.6\cdot0.025\cdot0.628318}}\cdot1.0183\\
=0.016361\;\frac{\mu b}{\text{rad}}\\
{R2_{TT}}^{s2,\alpha_{[p_{f}\pi^{+}][p\pi^{-}]}}_{\phi_{\pi^{-}}}&= {\frac{-220.61\cdot 1.70131}{\num{28.18e9}\cdot0.000117\cdot0.6\cdot0.025\cdot0.628318}}\cdot1.0183\\
=-0.014294\;\frac{\mu b}{\text{rad}}
\end{split}
\end{align}
}

\section{Introduction}
\label{sec:intro}

A key objective of the study of nucleon excited states ($N^*$) is to extract the resonance transition form factors that contain the information needed to explore the space-time scale-dependent description of the interaction of a virtual photon with the nucleon. These resonance transition form factors or $\gamma_vpN^*$ electrocouplings reveal the underlying quark-gluon degrees of freedom of these baryons and therefore serve to understand how their scale-dependent structure and mass emerge from QCD \cite{Achenbach:2025kfx,Carman:2023zke,Burkert:2025coj}.

Experimental observables from the charged single and double pion photo- and electroproduction channels serve as the primary source for the extraction of the electrocouplings for all prominent nucleon resonances over the two-dimensional (2D) landscape spanned by the kinematic variables $Q^2$ and $W$, which map out the space-time scale of the quark-gluon degrees of freedom and their strong interaction dynamics, as well as the mass range of the $N^*$ states, respectively.

While the combined studies of exclusive charged single and double pion production reveal the reliable extraction of the photo- and electrocouplings ($\gamma_{r,v}pN^*$), the combined studies of exclusive $\pi^+\pi^-p$ photo- and electroproduction off protons themselves represent an effective tool for the exploration of the spectrum and structure of excited nucleon states \cite{Burkert:2025coj,Mokeev:2022xfo}. The exclusive $\pi^+\pi^-p$ channel is sensitive to the contributions of most $N^*$s, particularly those within and above the third resonance region, which allows for the determination of their photo- and electrocouplings at different photon virtualities $Q^2$, providing for insight into the structure of the different $N^*$ states each with their own distinctive structural features.

The combination of the Jefferson Laboratory (JLab) electron beam and the large acceptance CLAS detector in Hall~B provide a unique opportunity to study the $\gamma_{r,v} p \to \pi^+\pi^- p'$ reactions with almost complete coverage of the final-state phase space. This capability is of particular importance for the extraction of the photo- and electrocouplings, as well as the hadronic decay widths into the $\pi\Delta$ and $\rho p$ final states. The $\pi^+\pi^- p$ electroproduction data from CLAS~\cite{Ri03,Fedotov:2008aa,Isupov:2017lnd,Fedotov:2018oan,Mokeev:2023zhq} provide the only available information on the nine independent single-differential and the fully integrated cross sections in the invariant mass range $W < 2.0$~GeV for $0.25~\mathrm{GeV}^2 < Q^2 < 5.0~\mathrm{GeV}^2$, and play a major role in the extraction of the electrocouplings of higher-lying $N^*$ states ($M >1.55$~GeV), e.g. the $\Delta(1600)3/2^+$, $\Delta(1620)1/2^-$, $\Delta(1700)3/2^-$, $N(1720)3/2^+$, and the new $N'(1720)3/2^+$ state seen in the combined studies of CLAS $\pi^+\pi^-p$ photo- and electroproduction data~\cite{Mokeev:2020hhu}. As of now, the electrocouplings of these states can only be determined from the data in the $\pi^+\pi^-p$ channel \cite{Mokeev:2015lda,Mokeev:2018zxt,Mokeev:2020hhu}, as the $\pi N$ channels do not have enough sensitivity. On the other hand, the electrocouplings of the $N(1675)5/2^-$, $N(1680)5/2^+$, and $N(1710)1/2^-$ were determined from analyses of the CLAS $\pi N$ electroproduction data~\cite{Aznauryan:2014xea,Park:2014yea} where this channel is more sensitive. 

Currently, the $\gamma_vpN^*$ electrocouplings are available from the CLAS data on exclusive $\pi N$, $\eta N$, $K\Lambda$, $K\Sigma$, and $\pi^+\pi^-p$ electroproduction for most $N^*$ states across the mass range up to 1.8~GeV at $Q^2$ from 0.2--5.0~GeV$^2$~\cite{Aznauryan:2011qj,Mokeev:2018zxt}, extended by the first results from global coupled-channel analyses of the electroproduction data from CLAS~\cite{Wang:2024byt,Kamano:2016bgm,Kamano:2018sfb}. A summary of the CLAS results on the $N^*$ electrocouplings can be found in Ref.~\cite{HillerBlin:2019jgp}.

Studies of the CLAS results on the $Q^2$-evolution of the $\gamma_vpN^*$ electrocouplings have considerably extended insight into the structure of $N^*$ states and the strong QCD dynamics underlying their generation. Investigations of the electrocouplings within continuum Schwinger methods (CSM) \cite{Segovia:2019jdk,Ding:2022ows} and quark model approaches \cite{Aznauryan:2012ec, Aznauryan:2012ba,Aznauryan:2018okk,Giannini:2015zia,Ramalho:2017pyc,Ramalho:2018wal,Ramalho:2023hqd}, in combination with global multi-channel analysis results \cite{Kamano:2016bgm,Kamano:2018sfb,Suzuki:2010yn}, have revealed the structure of $N^*$ states as a complex interplay between an inner core of three dressed quarks and an external meson-baryon cloud \cite{Aznauryan:2014xea,Burkert:2019bhp,Mokeev:2015lda,Mokeev:2025hhe}. This has resolved a long-standing puzzle on the role of the meson-baryon and quark degrees of freedom in the structure of these states.

The first CLAS data on the nine independent one-fold differential $\gamma_v p \to \pi^+\pi^-p'$ cross sections for $W$ from 1.4-2.0~GeV and $Q^2$ from 2.0-5.0~GeV$^2$ were published in 2017 \cite{Isupov:2017lnd}. In this paper we report updated cross sections based on the same dataset, but with improved statistical and systematic uncertainties, due to updates in the criteria for the exclusive event selection, more realistic event generators for the efficiency evaluations, and optimized cross section estimates within the inefficient (or blinded) areas of the CLAS detector compared to the work of Ref.~\cite{Isupov:2017lnd}. For the first time, we have been able to unambiguously show that we have achieved sufficient Monte Carlo (MC) simulation statistics necessary for a reliable evaluation of the event detection efficiencies across the broad seven-dimensional (7D) phase space of roughly 20 million cells. 

Furthermore, we report here the first results on the $TT$ and $LT$ components of the $\gamma_v p \to \pi^+\pi^-p'$ differential cross sections. These terms offer insight into the interference between different production amplitudes, which is of particular value to further refine amplitude analyses aimed at deducing the $\gamma_vpN^*$ electrocouplings from $\pi^+\pi^-p$ electroproduction data. A high-level analysis paper based on these cross section data to extract the $\gamma_vpN^*$ electrocouplings has already been prepared~\cite{2pi-hilevel}, and in the near future, the electrocouplings of all prominent $N^*$ states for $W$ up to 1.8~GeV and $Q^2$ from 2.0--5.0~GeV$^2$ will become available. 

The organization for the remainder of this paper is as follows. In Section~\ref{sec:formalism} the cross sections extracted from this analysis of $\pi^+\pi^-p$ electroproduction data, including both the single-differential and the photon-polarization-dependent cross sections, are defined along with the relevant kinematic variables and details on their extraction. Section~\ref{sec:analysis} describes the experiment, event selection procedures, data binning, yield extraction method, acceptance and efficiency corrections, and details the systematic uncertainty analysis. Section~\ref{sec:results} presents the cross sections and compares them to the available JLab-Moscow State University (JM) meson-baryon reaction model to provide insight into the resonant and non-resonant contributions as a function of kinematics. Finally, a summary of this work and our conclusions are given in Section~\ref{sec:conclusions}.
\section{Cross Section Formalism}
\label{sec:formalism}

\begin{figure}[htbp]
\begin{center}
\includegraphics[width=1.0\columnwidth]{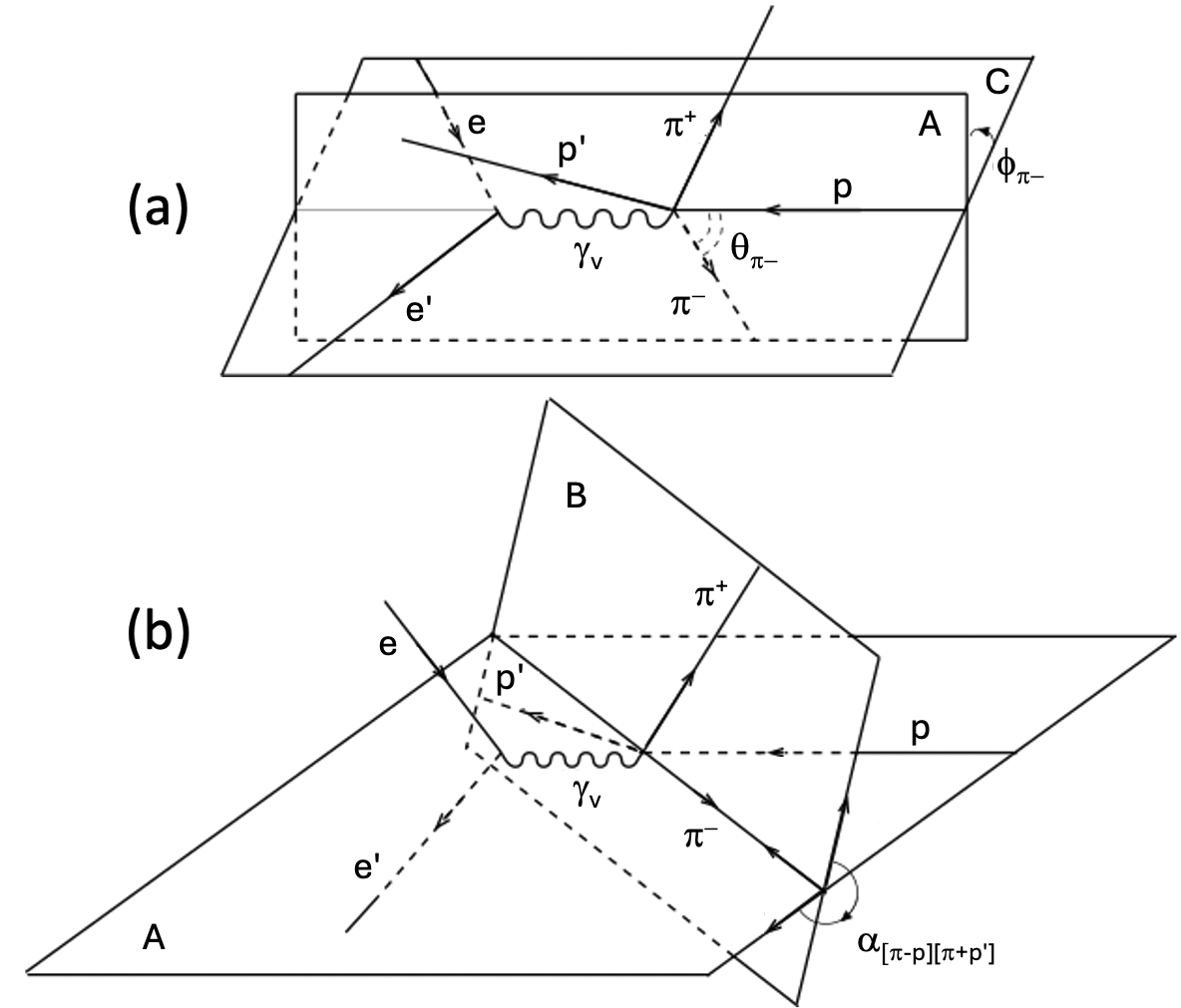}
\caption{Kinematic variables for the description of the reaction $\gamma_v p \to  \pi^+ \pi^- p'$ in the CM frame of the final state hadrons. Panel (a) shows the $\pi^-$ polar and azimuthal angles $\theta_{\pi^-}$ and $\phi_{\pi^-}$. Plane C represents the electron scattering plane. The $z$-axis is directed along the $\gamma_v$, while the $x$-axis is located in the electron scattering plane C and the $y$-axis forms a right-handed coordinate system. Plane A is defined by the three-momenta of the initial state proton and the final state $\pi^-$. Panel (b) shows the angle $\alpha_{[\pi^-p][\pi^+p']}$ between the two hadronic planes A and B or the plane B rotation angle around the axis aligned along the three-momentum of the final state $\pi^-$. Plane B is defined by the three-momenta of the final state $\pi^+$ and $p'$.} 
\label{fig:kinematics}
\end{center}
\end{figure}

For the $\gamma_v p \to \pi^+\pi^-p'$ reaction, the invariant mass of the final-state hadrons $W$ and the photon virtuality $Q^2$ unambiguously determine the initial-state virtual photon and proton four-momenta in their center-of-mass (CM) frame, with the $z$-axis directed along the three-momentum of the virtual photon, as illustrated in Fig.~\ref{fig:kinematics}(a). The final $\pi^+\pi^-p$ state is described by twelve variables, corresponding to the four-momenta of the three final-state hadrons. Energy-momentum conservation and the on-shell conditions for the final-state hadrons reduce the number of independent variables to five. Thus, for a given $W$ and $Q^2$, the reaction is fully described by the five-fold differential cross section $d^5\sigma/d^5\tau$, where $d^5\tau$ denotes the differential in these five independent kinematic variables. There are three possible choices for these five variables~\cite{Byckling:1971vca}

\begin{equation}
\label{set_var}
\begin{aligned}
1.\,& M_{p'\pi^+},\, M_{\pi^+\pi^-},\, \theta_{\pi^-},\, \phi_{\pi^-},\, 
\alpha_{[\pi^-p][\pi^+p']} \\
2.\,& M_{p'\pi^+},\, M_{p'\pi^-},\, \theta_{\pi^+},\, \phi_{\pi^+},\, 
\alpha_{[\pi^+p][\pi^-p']}\\
3.\,& M_{p'\pi^+},\, M_{\pi^+\pi^-},\, \theta_{p'},\, \phi_{p'},\, 
\alpha_{[\pi^+\pi^-][pp']}.
\end{aligned}
\end{equation}

The sets include the 3 di-hadron invariant mass combinations ($M_{p'\pi^+}$, $M_{\pi^+\pi^-}$, $M_{p'\pi^-}$), the 3 hadron CM polar angles ($\theta_{\pi^-}$, $\theta_{\pi^+}$, $\theta_{p'}$), the 3 hadron CM azimuthal angles ($\phi_{\pi^-}$, $\phi_{\pi^+}$, $\phi_{p'}$), defined as the angle between the electron scattering plane and the planes formed by the three-momenta of the final-state hadrons and the target proton in the CM frame, and the 3 inter-hadron plane CM $\alpha$ angles, where, for example, $\alpha_{[\pi^-p][\pi^+p']}$ is the angle between plane A formed by the $\pi^-$ and target proton $p$ and plane B formed by the $\pi^+$ and scattered proton $p'$ (see Fig.~\ref{fig:kinematics}(b)). The angles $\alpha_{[\pi^+p][\pi^-p']}$ and $\alpha_{[\pi^+\pi^-][pp']}$ are defined analogously. More details are included in Ref.~\cite{Iulia-paper}. These final state variables are collectively referred to as $X_{ij}$, where the index $i = 1 \to 3$ represents the variable set and the index $j = 1 \to 5$ represents the variable within the set. It is insightful to think of the $X_{ij}$ forming a five-dimensional (5D) kinematic phase space within each ($Q^2,W)$ bin, where the differential over the final state kinematic variables in the 5D cell is denoted by $d^5\tau$.

The 5D differential cross section of the reaction $\gamma_v p \to \pi^+\pi^-p'$ in a given $(Q^2,W)$ bin can be factored from the full 7D cross section using the virtual photon flux factor $\Gamma_v$ that separates, under the single-photon exchange approximation, the electron scattering part described within quantum electrodynamics from the hadronic part that provides information on the $N^*$ structure as

\begin{equation} 
\label{eqn5Dxsec}
\frac{d^5\sigma}{d^5\tau}\left(\Delta Q^2,\Delta W\right) = \frac{1}{\Gamma_v} \frac{d^7\sigma}{dW dQ^2 d^5\tau},
\end{equation}

\noindent
where $d^5\tau$ is the bin volume in the 5D space and the virtual photon flux factor is defined as~\cite{Knochlein:1995qz}
\begin{equation}
\label{eq:flux}
\Gamma_v = \frac{\alpha}{4\pi}\frac{W}{M_p^2 E^2}\frac{W^2-M_p^2}{Q^2} \frac{1}{1-\epsilon}.
\end{equation}
\noindent
Here $\epsilon$ is the degree of the transverse virtual photon polarization defined in terms of the virtual photon energy $\nu$, the electron beam energy $E$, and the electron scattering angle $\theta_{e'}$ in the laboratory frame as
\begin{equation}
\epsilon=\left(1+2\frac{\nu^2}{Q^2}\tan^2{\frac{\theta_{e'}}{2}} \right)^{-1}.
\end{equation}

Further details on the cross section formalism for charged double-pion electroproduction are provided in Ref.~\cite{Iulia-paper}. For the case where there is no beam and target polarization, the cross section in the CM frame can be written as
\begin{multline}
\label{d5t}
\frac{d^5\sigma}{d^5\tau} = \sigma_T + \epsilon_L \sigma_L + \epsilon( {{^c}\sigma_{TT}} \cos 2 \phi + {{^s}\sigma_{TT}} \sin 2 \phi)\, + \\
\sqrt{2 \epsilon_L (1 + \epsilon)}({{^c}\sigma_{LT}} \cos \phi + {{^s}\sigma_{LT}} \sin \phi).
\end{multline}
\noindent
Here $\epsilon_L = (Q^2/\nu^2) \cdot\epsilon$, typically referred to as the longitudinal polarization in the laboratory frame, is kinematically related to the transverse polarization of the virtual photon, $\epsilon$. This form demonstrates that the differential electroproduction cross section includes both $\phi$-independent and $\phi$-dependent components. The $\phi$-independent terms are defined by electroproduction amplitudes associated with purely transverse or longitudinal virtual photons, whereas the $\phi$-dependent terms are defined by interference of the transverse and longitudinal polarization amplitudes of the virtual photons in mixed helicity states.

The terms $\sigma_T$, $\sigma_L$, $^c\sigma_{TT}$, $^s\sigma_{TT}$, $^c\sigma_{LT}$, and $^s\sigma_{LT}$ are known as structure functions, and encompass, in analogy to single-pion electroproduction, the full information on the $\pi^+\pi^-p$ electroproduction cross sections. The $c$ and $s$ superscripts denote their dependence on the $\cos \phi$/$\cos 2\phi$ terms or the $\sin \phi$/$\sin 2\phi$ terms, respectively, where the $\sin \phi$ and $\sin 2\phi$ terms only appear in double-pion electroproduction and are absent in unpolarized beam/target single-meson electroproduction. They depend on the variables $Q^2$, $W$, and all $d^5\tau$ variables of the final state, with the exception of $\phi$. The $\phi$-dependencies of the different terms allow for their separate extraction via a straightforward fit of the $\phi$ dependence of the cross section. If Eq.~\ref{d5t} is integrated over $\phi$, all of the interference contributions ($^c\sigma_{TT}$, $^s\sigma_{TT}$, $^c\sigma_{LT}$, and $^s\sigma_{LT}$) vanish, and only the unpolarized cross section terms ($\sigma_T$ and $\sigma_L$) remain.

\subsection{Single-Differential Observables} 
\label{sec:obs_1D}

The charged double-pion electroproduction data provided from CLAS to date have focused on measurements of the $\phi$-integrated single-differential cross sections \cite{Isupov:2017lnd,Fedotov:2018oan}. The reason for this is that the statistics in the 7D bins are much too small, demanding the integration over four of the five hadronic cross section dimensions to allow for further analysis by fitting these single-differential cross sections \cite{Mokeev:2015lda,Mokeev:2018zxt,Mokeev:2020hhu}. The corresponding observables were directly obtained by projecting the 5D differential cross sections in each $(Q^2,W)$ bin onto each dimension of the $d^5\tau$ variables. As introduced above, these 5D cross sections are obtained in 3 variable sets with 5 variables in each. Therefore, there are a possible total of 15 single-differential cross sections, although not all of them are unique. The di-hadron mass combinations of the particles in the final state are repeated in the variable sets, which reduces the 15 possible combinations to 12. Additionally, as the data have been integrated over $\phi$, the total number of observables is reduced to 9. 

The 9 variables used for obtaining the single-differential cross sections in each $(Q^2,W)$ bin include

\begin{enumerate}
	\item $M_{p'\pi^+},\theta_{\pi^-},\alpha_{[\pi^-p][\pi^+p']}$
    \item $M_{p'\pi^-},\theta_{\pi^+},\alpha_{[\pi^+p][\pi^-p']}$
    \item $M_{\pi^+\pi^-},\theta_{p'},\alpha_{[\pi^+\pi^-][pp']}$.
\end{enumerate}

\subsection{Photon-Polarization-Dependent Cross Sections} 
\label{sec:obs_R2}

To access the $\phi$--dependent parts of the photon-polarization-dependent cross section terms of Eq.~\ref{d5t}, the 5D cross sections are projected as $d^2\sigma/dX_{ij}d\phi_i$. Generalizing the form given in Eq.~\ref{d5t}, we can express these 2D projections as

\begin{align} 
\label{eqn2Dphiproj}
\frac{d^2\sigma}{dX_{ij}d\phi_i}=
\begin{split}
&{R2_T}^{X_{ij}}_{\phi_i} + {R2_L}^{X_{ij}}_{\phi_i} + \\ 
&{R2_{LT}}^{c,X_{ij}}_{\phi_i}\cos\phi_i + {R2_{TT}}^{c,X_{ij}}_{\phi_i}\cos2\phi_i + \\
&\delta_{X_{ij}\alpha_i}\left({R2_{LT}}^{s,\alpha_i}_{\phi_i}\sin\phi_i + {R2_{TT}}^{s,\alpha_i}_{\phi_i}\sin2\phi_i\right),
\end{split}
\end{align}
\noindent
where $\phi_i$ is the $\phi$ angle from the $i^{th}$ variable set with $X_{ij} \neq \phi_i$. Our notation here for the contributions from the virtual-photon-polarization independent/dependent components of the two-fold differential cross sections is based on the single-photon exchange formalism applicable to any exclusive meson electroproduction channel. Considering the single-meson electroproduction formalism of Ref.~\cite{Knochlein:1995qz}, we replace ``R'' with ``R2'' for the notation of the two-pion channel.  

{In contrast to the polarization observables for single meson electroproduction listed in Table I of Ref.~\cite{Knochlein:1995qz}, the presence of an additional particle in the final state adds \RDwOO{} and \REwOO{} to the list of observables. The ``00'' superscript refers to the fact that both the beam and target are unpolarized in this analysis, but to simplify the notation, this superscript is generally omitted in this paper. Note that owing to parity conservation, the coefficients of the sine functions (\RD{} and \RE{}) are non-zero only when $X_{ij} = \alpha_i$, where $\alpha_i$ is the respective $\alpha$ angle in the $i^{th}$ variable set. This is accounted for by the delta function term $\delta_{X_{ij}\alpha_i}$.}

To extract the R2 observables from the 5D cross sections, $\phi_i$ projections were made for data corresponding to each of the $X_{ij}$ bins. The $\phi$ dependence of these projections was then fitted using the sinusoidal functional form of the $\phi$ dependence of Eq.~\ref{eqn2Dphiproj}. These photon-polarization-dependent cross sections have been extracted for the first time in this analysis.
  
As compared to the 9 single-differential cross sections discussed in Section~\ref{sec:obs_1D}, the total number of polarization observables measured through the 2D projections is significantly increased. This is because there are 5 polarization observables (\RA, \RB, \RC, \RD{}, and \RE{}) and each of them is extracted for all relevant $X_{ij}$. Therefore, there are 42 possible polarization observables given by
\begin{itemize}
	\item 36 from the possible \RA, \RB, \RC: 3 observables $\times$ 3 variable sets $\times$ 4 variables per variable set ($X_{ij} \neq \phi_i$) and
	\item 6 from the possible \RD{} and \RE{}: 2 observables $\times$ 3 variable sets $\times$ 1 variable per variable set ($X_{ij} = \alpha_i$ only).
\end{itemize}
\indent
However, of these 42 observables, only 30 provide new information as the 12 related to \RA are equivalent to the 9 single-differential cross sections of Section~\ref{sec:obs_1D}. The relevant $X_{ij}$ for each of the photon-polarization-dependent terms are given by (where only $\alpha_i$ are valid for \RD{} and \RE{})
\vspace{2mm}
\begin{enumerate}
	\item $M_{p'\pi^+},M_{\pi^+\pi^-},\theta_{\pi^-},\alpha_{[\pi^-p][\pi^+p']}$
    \item $M_{p'\pi^+},M_{p'\pi^-},\theta_{\pi^+},\alpha_{[\pi^+p][\pi^-p']}$
    \item $M_{p'\pi^+},M_{\pi^+\pi^-},\theta_{p'},\alpha_{[\pi^+\pi^-][pp']}$.
\end{enumerate}

\subsection{Extraction of Observables from Experimental Data}
\label{sec:extraction}

The analysis of the 7D cross sections is based on the experimental two-pion data yields, constants, and correction factors, see Eq.~\ref{eqn7Dxsec}, described in detail in Section~\ref{sec:analysis}. Practically, this extraction employed a 7D sparse histogram architecture and, hence, the approach described here directly references to the \textit{bins} of such histograms, which are basically the data points of the analysis. The binning of the histogram is setup to optimally extract cross sections given not only the resolution of the CLAS detector, but also the intrinsic width of the $N^*$ resonances, both of which affect the final choice for the $W$ bin width. 

The form of the 7D cross section used to extract the cross sections is expressed as

\begin{widetext}
\begin{equation} 
\label{eqn7Dxsec}
\frac{d^7\sigma}{dW dQ^2 d^5\tau} = \frac{1}{{\cal L}}\frac{1}{R}\frac{\left(\frac{\Delta^7N_{\text{ER}}-\Delta^7N_{\text{etgt}}\cdot Q_{\text{ratio}}}{A\cdot \epsilon^{\text{CC}}}+\Delta^7N_{\text{EH}}\right)}{\Delta W \Delta Q^2 \Delta^5 \tau}.
\end{equation}
\end{widetext}
\noindent
Note that the left-hand side of this 7D cross section is written with differentials ``$d$" to simplify the notation. However, as the presented results are not bin-centered cross sections, but instead bin-averaged cross sections, we could have more properly written the functional replacing ``$d$" with ``$\Delta$". In this expression, $\Delta^7N_{\text{ER}}$ and $\Delta^7N_{\text{EH}}$ are the total number of \fullrctn{} events in an experimentally accessible and inaccessible 7D bin, respectively (see Sections~\ref{corr-yields}.2,3). The notion of an inaccessible bin, called a \textit{kinematical hole}, arises in the analysis as certain 5D bins in the simulation do not contain reconstructed two-pion events due to the limited detector acceptance. These bins with zero acceptance are called holes and were filled based on the event generator used to model the two-pion cross section (see Section~\ref{corr-yields} and Ref.~\cite{ananote} for details). $\Delta^{7}N_{\text{etgt}}$ are the \fullrctn{} events from empty target runs in an experimentally accessible 7D bin scaled by $Q_{\text{ratio}}$, the ratio of the integrated Faraday Cup charge from the production to empty target runs (see Section~\ref{corr-yields}.2). The term $\Delta W \Delta Q^2 \Delta^5\tau$ represents the 7D volume of a bin (see Section~\ref{binning}), $A$ is the acceptance correction factor determined using simulation (see Section~\ref{sec:acc}), and $\epsilon^{\text{CC}}$ is the efficiency factor for the electrons in the Cherenkov detector determined from data runs with the Cherenkov detector not included in the trigger. The final terms in Eq.~\ref{eqn7Dxsec} are the radiative correction factor $R$ (see Section~\ref{corr-yields}.1) and the integrated beam-target luminosity ${\cal L}$ (see Eq.~\ref{eqnLuminosity}).
\section{Experiment Description and Data Analysis}
\label{sec:analysis}

\subsection{Dataset and Detector Information}

The dataset employed for this analysis was collected in the period from Oct. 2001 to Jan. 2002 using the CLAS spectrometer in Hall~B~\cite{CLAS:2003umf}. An electron beam of energy 5.754~GeV was directed onto a 5-cm-long liquid-hydrogen target located in the center of the detector along the electron beamline.

The main magnetic field of CLAS was provided by six superconducting coils, which produced an approximately toroidal field in the azimuthal direction around the beam axis. The gaps between the cryostats, called sectors, were instrumented with identical detector packages. Each of the six sectors consisted of three sets of multi-layer drift chambers (DC) for charged particle tracking and momentum determination, Cherenkov counters (CC) for electron/pion separation, scintillator counters (SC) for charged particle timing, and electromagnetic calorimeters (EC) for electron and neutral particle identification. The polarity of the torus magnet was set to bend negatively charged particles in toward the electron beamline.

To reduce the electromagnetic background resulting from M{\o}ller scattering oﬀ atomic electrons in the target and the target cell, a small normal-conducting toroidal magnet (called the mini-torus) was placed symmetrically about the target inside of the first DC package to remove these low energy electrons from the detector acceptance. A Faraday cup was located at the end of the beamline to determine the integrated beam charge passing through the target. The polar angle coverage for electrons ranged from 8$^\circ$ to 45$^\circ$ and for charged hadrons ranged from 8$^\circ$ to 142$^\circ$. The resolution of the reconstructed polar and azimuthal angles was better than 2~mrad. The CLAS detector was designed to track charged particles with momenta greater than $\sim$200~MeV with a resolution $\Delta p/p$ in the range of 0.5 to 1\%.

The data were taken with typical electron beam currents of 5~nA at a luminosity of $1 \times 10^{34}$~cm$^{-2}$s$^{-1}$. The CLAS event readout was triggered by a coincidence between a Cherenkov segment hit that matched to a reconstructed cluster in the calorimeter in a single sector, generating a data acquisition event rate of $\sim$2~kHz with live times greater than 90\%. The large acceptance of CLAS enabled detection of the final-state electron, proton, and charged pions. In this section, details are provided on the procedures for particle identification, the cuts used to isolate the exclusive charged two-pion final state, and other cuts and corrections to convert the measured event yields into the final reported differential cross sections. An earlier analysis of the same dataset provides additional information~\cite{Isupov:2017lnd}.

\subsection{Electron Identification}
\label{sctn-EID}

The initial electron candidate requirement was based on selection of tracks that satisfied the electron hardware trigger definition. These negatively charged particles (defined by their curvature in toward the electron beamline) were required to have a valid track status from reconstruction. It was further required that their vertex trace-back position to the electron beamline was within the range of the target location of $[-8.0,0.75]$~cm (accounting for tracking resolution), thus removing any electrons that were scattered off the target insulation layers or the foam scattering chamber exit window that was located 2~cm downstream of the target. 

Additional cuts were then applied using the CC and EC, to further refine the final electron sample from the candidate tracks. Each electron candidate was required to register a signal in the CC above a threshold defined in terms of the recorded number of photoelectrons, have a reconstructed momentum above the calorimeter trigger energy threshold of $>0.70$~GeV, a minimum energy deposited in the EC $>0.06$~GeV to remove minimum-ionizing pions, and be reconstructed within the defined EC detector fiducial volume. 

\begin{figure}[htb]
  \centering
  \includegraphics[width=0.4\textwidth]{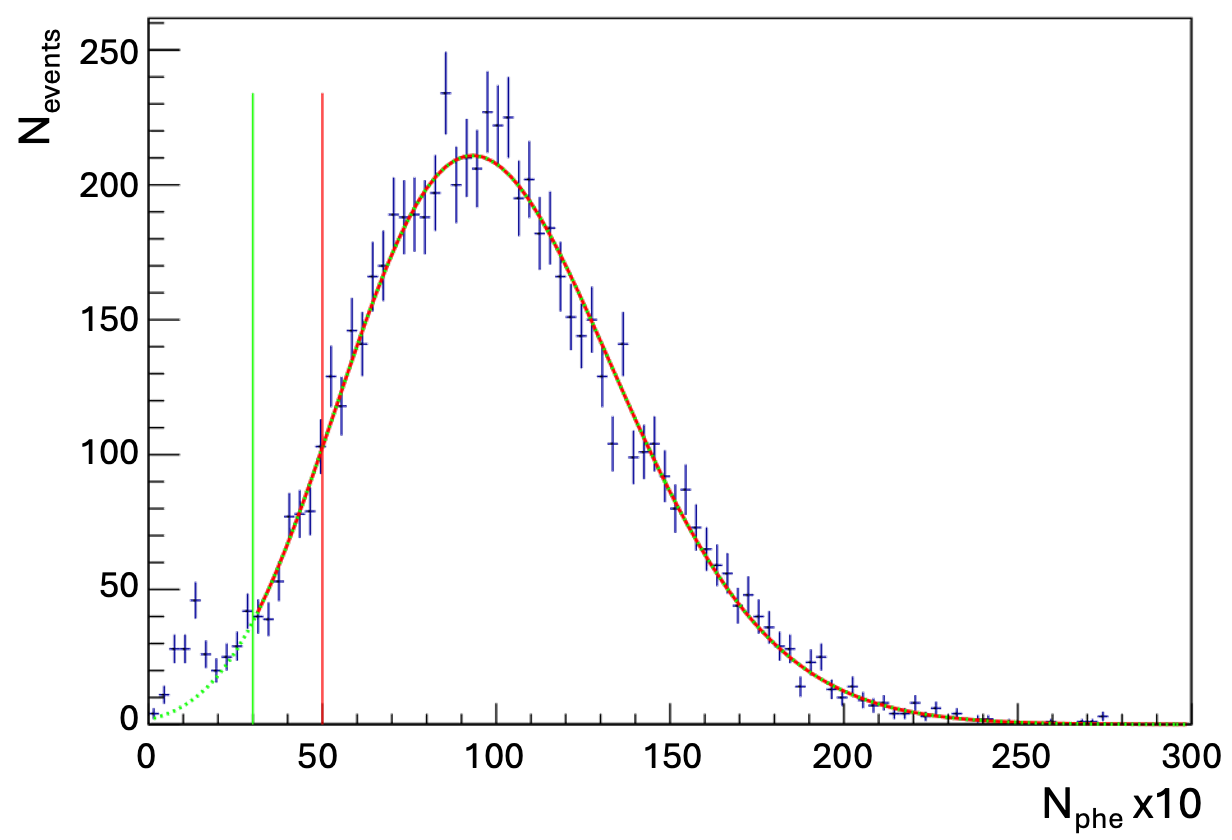}
  \caption{Distribution of the number of photoelectrons for a representative photomultiplier tube (PMT) within the CC system. The green and red dotted curves show the Poisson fits to the distribution. The solid green and red lines represent the loose and tight thresholds of this cut, respectively.}
  \label{fignphe}
\end{figure}

Figure~\ref{fignphe} shows the photoelectron distribution for a representative PMT in the CC system. While the photoelectron distribution due to electron tracks defines a Poisson distribution, a small contamination due to pions and PMT noise is present at low photoelectron values. A cut on the minimum number of CC photoelectrons $nphe_\text{min}$ was applied to the data to remove this peak near 0. The ratio of the Poisson distribution integrals obtained for the number of photoelectrons in the range from $[nphe_\text{min},50]$ to $[0,50]$ was used to determine the CC efficiency correction for each CC element. The nominal minimum photoelectron cut was set to 5. The procedure was applied to both the data and MC. Loose and tight cut values for this threshold were employed to study the systematic uncertainty associated with this cut. For extracting the observables, the loose cut was used and the difference between the two cuts affects the final extracted cross sections at the level of 1\%.

\begin{figure}[htb]
  \centering
  \includegraphics[width=0.8\columnwidth]{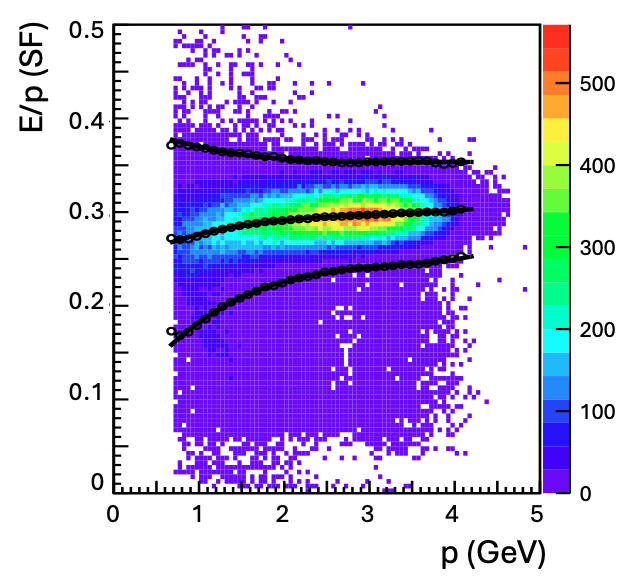}
  \caption{Distribution of the sampling fraction $E_{\text{tot}}/p$ for the electron candidates in a representative CLAS sector. The top and bottom black lines represent the $\pm 3\sigma$ cut limits about the central mean value as a function of the electron momentum.}
  \label{figSFvp_ER}
\end{figure}

Finally, the EC for CLAS was based on a lead-scintillator sampling calorimeter design. A sampling fraction (SF) cut was applied to select the electron candidates and remove any remaining $\pi^-$ contamination. For the EC the nominal SF was roughly 30\% with a slight dependence on the deposited energy. The SF is defined as the ratio of the total measured cluster energy $E_{\text{tot}}$ relative to the track momentum as reconstructed by the DC, $SF = E_{\text{tot}}/p$. Due to the inherent statistical nature of the energy deposition process and the resolution of the detector, the SF is Gaussian distributed. To develop an effective cut selection on the SF distribution for each CLAS sector, the SF distribution was fit with a third-order polynomial to determine the mean. The distribution was then fit with a Gaussian as a function of momentum to determine the $\pm 3\sigma$ limits that defined the SF cut. Figure~\ref{figSFvp_ER} shows a representative SF vs. $p$ distribution for the electron candidates in a single CLAS sector showing the cut definition relative to the fitted mean.

\subsection{Charged Hadron Identification}

In this analysis only the $p$ and $\pi^+$ were required to be measured in the detector. All positively charged hadrons that registered a signal in the DC and SC were considered to be candidates for the $p$ and $\pi^+$ in each event. The charged hadrons were required to have a DC track that matched to a hit in the SC. The tracks were required to have their curvature consistent with that for a positively charged particle and have a minimum momentum of 250~MeV. 

\begin{figure}[htb]
\centering
 \includegraphics[width=0.9\columnwidth]{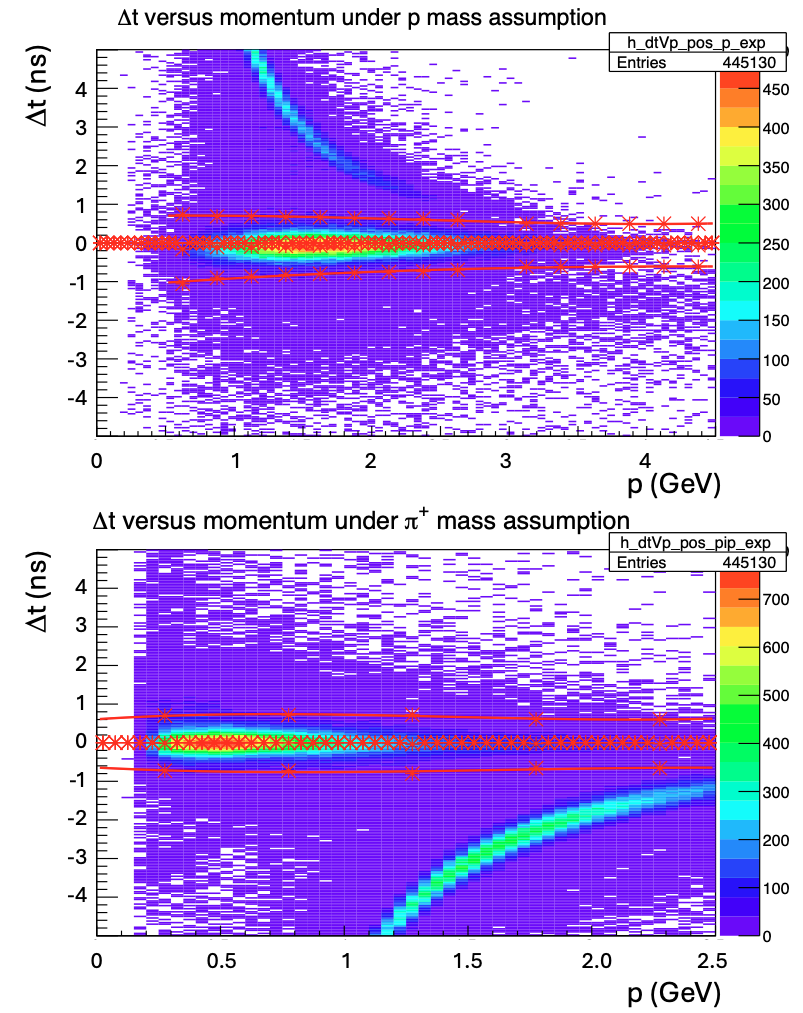}
 \caption{Distributions of the $\Delta t$ cut (ns) for protons (top) and positive pions (bottom) as a function of hadron momentum. The region between the red lines on each plot represents the selected $\Delta t$ cut limits.}
 \label{figdt_ER} 
\end{figure}

Hadron identification was performed using a timing cut. The timing quantity of interest ($\Delta t = t_1 -t_2$) was the  difference in the time ($t_1$) for a particle to travel the path $d$ from its production vertex to the SC system and the time ($t_2$) for an assumed particle mass in conjunction with the measured momentum $p$. The time $t_2$ (using natural units with $c$=1) is given by

\begin{equation}
t_2 = \frac{d}{\beta_2},
\end{equation}

\noindent
where $\beta_2$ is given by

\begin{equation}
\beta_2 = \frac{p}{\sqrt{m_2^2 + p^2}},
\end{equation}

\noindent
with $m_2$ is the assumed particle mass. The time $t_2$ was computed for all positively charged particles assuming the mass of the pion, kaon, and proton. The mass assumption that minimized $\Delta t$ was assigned as the particle type. The time difference $\Delta t$ is given by

\begin{equation}
\Delta t = t_1 \left( 1 - \sqrt{ \frac{p^2 + m_2^2}{p^2 + m_1^2}} \right ),
\end{equation}

\noindent
where $m_1=p/\beta\gamma$ is the computed particle mass. Figure~\ref{figdt_ER} shows plots of $\Delta t$ vs. $p$ for the positive hadrons under the proton (top) and pion mass assumptions (bottom). Overlaid on these plots are the cut lines (red) to select the hadrons. These limits are based on a $\pm 3\sigma$ cut about the band at $\Delta t = 0$.

\subsection{Fiducial Cuts}
\label{sec:fid}

\begin{figure}[htb]
  \centering
  \includegraphics[width=0.8\columnwidth]{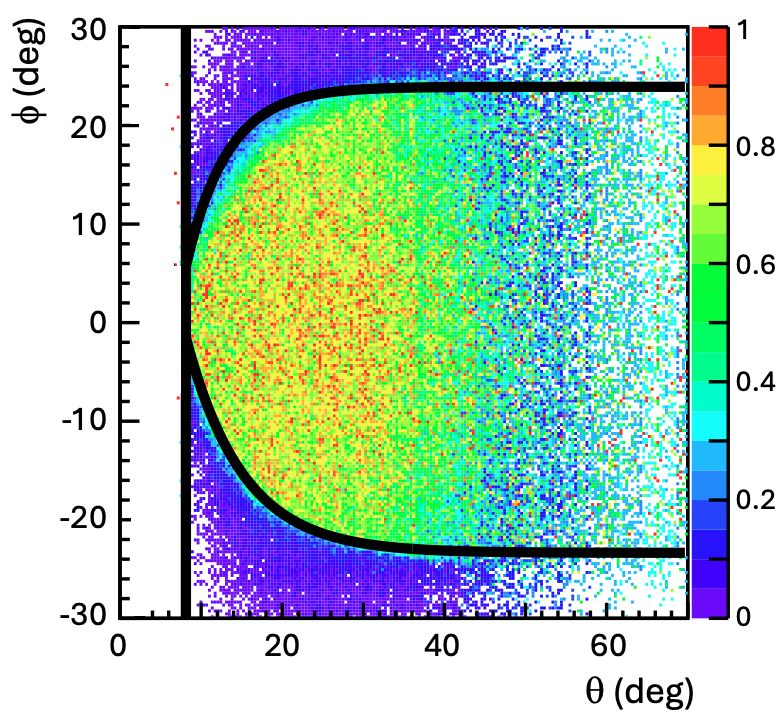}
  \caption{Illustration of the drift chamber based fiducial cuts for protons for a representative sector in terms of azimuthal angle vs. polar angle. The vertical line at $\theta = 8^\circ$ represents the minimum angle cut.}
  \label{dc-fiducial}
\end{figure}

The CLAS detector had incomplete azimuthal angle acceptance due to the mini-torus coils, the torus cryostats, and the edges of the drift chamber volumes. In addition, the CLAS detector acceptance had limits due to the minimum and maximum polar angle coverage of the detector subsystems. To address this reality, fiducial cuts were defined for the event selection that were matched between data and simulation. These cuts served to define a precise geometrical region of CLAS where the detection efficiency was reasonably large and uniform. The cuts for both electrons and positively charged hadrons depend on the track momentum and scattering angles. The fiducial regions were defined by placing cuts on the azimuthal vs. polar angle distributions as shown in Fig.~\ref{dc-fiducial}.

In addition, for accuracy in the determination of the electron energy, it was important that the shower in the EC did not occur too close to the edge of the calorimeter where a portion of the shower energy could escape from the detector volume. To ensure that the shower was fully contained within the EC, the electron hit position on the face of the EC was determined by extrapolating the electron track beyond the outermost DC layer. It was then required that all cluster centroids were at least 10~cm away from the three faces of the triangular EC modules.

In this experiment, some regions of the CLAS detector were inefficient, because of localized holes in the DC (due to broken wires or failed amplifiers) or bad SC paddle PMTs. The inefficient detector regions were identified in plots of the measured track momentum $p$ vs. polar angle $\theta$. These regions were cut out of both the data and MC samples, providing a good match between the data and simulated detector acceptances (see Ref.~\cite{Isupov:2017lnd} for more details).

\subsection{Event Selection} 
\label{sec:event-sel}

The final event sample selection used the computed missing mass to identify events that belong to the exclusive $ep \to e'\pi^+\pi^-p'$ reaction channel. Most of the $\pi^-$ from this reaction were bent out of the acceptance of the CLAS detector so only the final state $e$, $p$, and $\pi^+$ were required to be reconstructed. The squared missing mass in the exclusive charged two-pion electroproduction reaction from a proton target is given by

\begin{equation}
  MM^2=(p_{\gamma_v}^\mu + p_p^\mu - p_{p'}^\mu - p_{\pi^+}^\mu) \cdot (p_{\gamma_v \mu} + p_{p \mu} - p_{p' \mu} - p_{\pi^+ \mu}),
\end{equation}
\noindent
where $p_{\gamma_v}$, $p_p$, $p_{p'}$ and $p_{\pi^+}$ are the four-momenta of the virtual photon, target proton, final state proton, and final state positive pion, respectively. For the events of interest, $MM^2$ is equal to the squared mass of the $\pi^-$ smeared by the resolution function of the detector. 

\begin{figure*} 
  \includegraphics[width=1.00\textwidth]{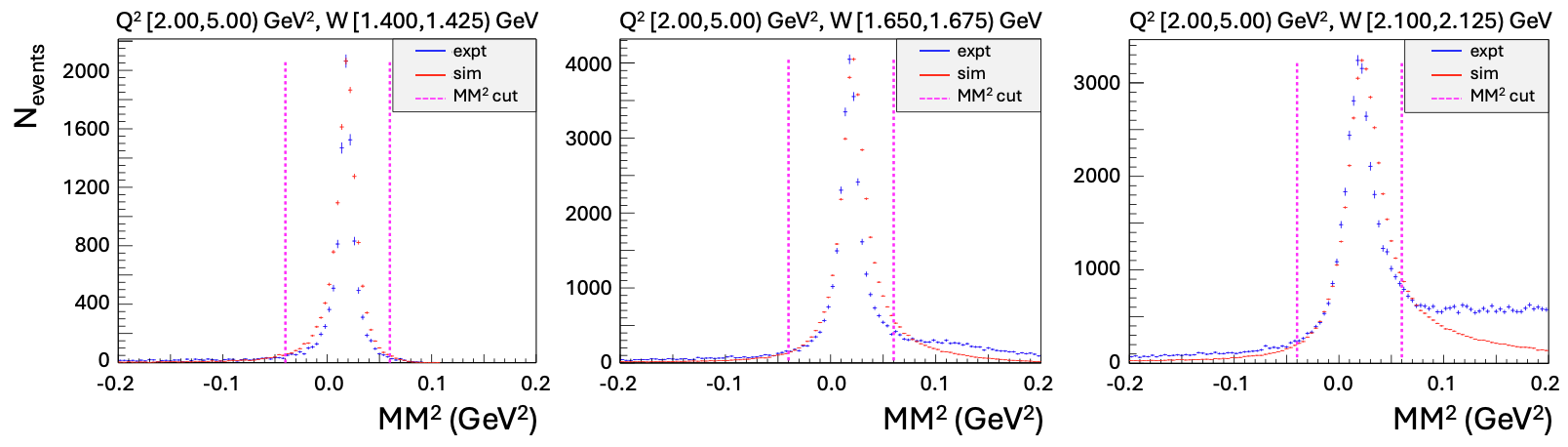}    
  \caption{$Q^2$ integrated $MM^2$ distributions from data (blue) and MC (red) in $W$ bins from the low, middle, and high $W$ ranges of the analysis as labeled for $Q^2$ from $2.0$ to $5.0$~GeV$^2$. The vertical dashed magenta lines represent the nominal $MM^2$ cut employed from $-0.04$ to $0.06$~GeV$^2$.}
  \label{fig_evtsel_MM2}
\end{figure*}

Figure~\ref{fig_evtsel_MM2} shows the $Q^2$ integrated $MM^2$ distributions from data (blue) and MC (red) in $W$ bins from the low, middle, and high $W$ range of the analysis: [1.400,1.425]~GeV, [1.650,1.675]~GeV, and [2.100,2.125]~GeV. The nominal $MM^2$ cut was defined in the range from $-0.04$ to 0.06~GeV$^2$ and was chosen to optimize the  signal-to-background ratio. The simulated momentum, energy, and timing resolutions were smeared to match the data so the same analysis cuts were employed for both data and MC event analysis. 

\begin{figure*} 
  \includegraphics[width=1.00\textwidth]{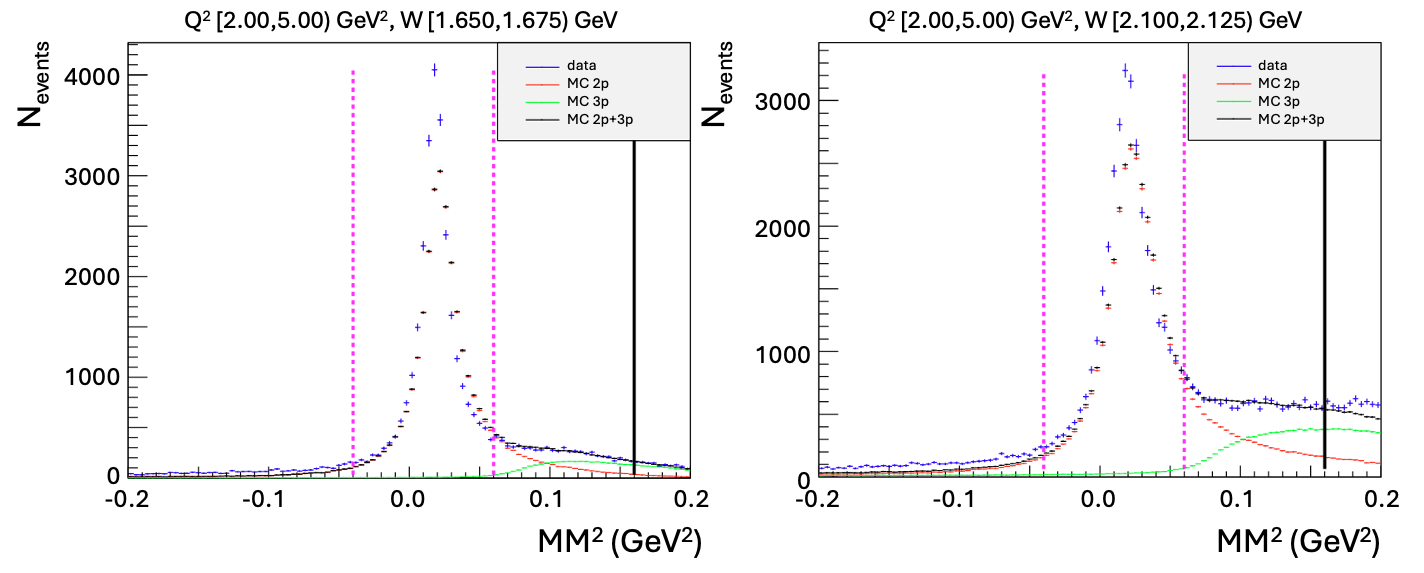}    
  \caption{Example $MM^2$ distributions from data (blue) compared to the normalized $2\pi$ (red) and $3\pi$ (green) contributions from MC for two representative $W$ bins as labeled. The vertical dashed magenta lines represent the nominal $MM^2$ cut. The black points are the sum of the MC $2\pi$ and $3\pi$ contributions. The red $2\pi$ points show the strength of the radiative tail contributions on the right-side of the $\pi^-$ peak. The black vertical line marks the $4\pi$ threshold.}
  \label{fig_mm2bck}
\end{figure*}

The main source of background in the $MM^2$ cut region was due to contributions from the $ep \to e'\pi^+\pi^-\pi^0p'$ channel. The procedure to quantify and account for this background was to simulate the $3\pi$ channel employing a phase space event generator and using the same cuts and corrections as for the $2\pi$ analysis. For each $W$ bin, the leakage of the $3\pi$ channel into the $MM^2$ cut limits for the $2\pi$ sample was determined. The contribution was found to increase with $W$ from 0\% at $W$=1.400~GeV up to 2.5\% at $W$=2.125~GeV. A correction factor was applied to the determined $2\pi$ event yields. The systematic uncertainty of the correction was estimated to be 1\% on the cross sections. Figure~\ref{fig_mm2bck} shows the background contributions within the $MM^2$ data range from the $3\pi$ reaction for two representative $W$ bins.

\subsection{Corrected Event Yields}
\label{corr-yields}

In this analysis the exclusive $ep \to e'\pi^+\pi^-p'$ reaction was selected using the four-momenta of the reconstructed scattered electron and the two positively charged hadrons $p$ and $\pi^+$. The remaining exclusive $\pi^-$ was tagged using the missing mass technique. The analysis determined the total number of measured events, $\Delta^7N_{\text{ER}}$, within each 7D kinematic bin defined by $Q^2$, $W$, and $\tau^5$ (see Section~\ref{binning} for details on the kinematic binning). In order to select the $ep \to e'\pi^+\pi^-p'$ events from all possible events that can result from the interaction of the electron with the target proton and remove any regions of the detector that were inefficient or non-functional, several selection criteria were applied. While these criteria kept most of the reaction events, some were lost in the process. These lost events were accounted for with an acceptance correction factor for each kinematic bin that was determined from the CLAS GEANT3 MC simulation (see Section~\ref{sec:acc} for details).

The cross section expression in Eq.~\ref{eqn7Dxsec} includes three correction terms, which are detailed in Ref.~\cite{ananote}, as follows:

\begin{enumerate}
\item Radiative correction factor $R$. Radiative eﬀects must be taken into account when determining an electroproduction cross section. To calculate the acceptance a MC event generator was employed that contained radiative eﬀects, including photon emission from the initial and scattered electron, as well as soft loop corrections. The acceptance-corrected yield in each kinematic bin was divided by the corresponding radiative correction factor as shown in Eq.~\ref{eqn7Dxsec}. The calculation of $R$ relied on the ``peaking approximation" to compute the radiative eﬀects \cite{Mo:1968cg}. Unlike complete next-to-leading order corrections, in this approach, photons were emitted only along the direction of the initial or final state electrons. This approximation reduces the size of the matrix element of the real photon emission part and simplifies the integration over the emitted photon phase space. In addition, emission of soft photons was summed by exponentiation. The radiative correction factor was applied to the binned cross sections defined as the ratio of the integrated binned cross sections for a process with radiation on to that with radiation off. Given that the hadronic vertex is not considered in this correction, the radiative correction employed depends only on $Q^2$ and $W$ and is given by

\begin{equation}
R = \frac{d^2\sigma_{\text{rad}}(Q^2,W)}{d^2\sigma_{\text{no-rad}}(Q^2,W)}.
\end{equation}

\item Empty-target correction factor $\Delta^7 N_{\text{etgt}}\cdot Q_{\text{ratio}}$. The event selection includes a target vertex cut that included the $15$-$\mu$m-thick target cell aluminum endcap windows. To account for these events and remove them from the final yield, data from runs taken with an empty cryocell were employed scaling the collected data to account for the relative luminosity difference. The events after the empty target contribution subtraction are given by

\begin{equation}
N_{\text{hydrogen}} = N_{\text{ER}} - N_{\text{etgt}} \cdot Q_{\text{ratio}},
\label{ETcorrectedYield}
\end{equation}

\noindent
where $N_{\text{hydrogen}}$, $N_{\text{ER}}$, and $N_{\text{etgt}}$ are the events from hydrogen, the full target cell including endcap windows, and the empty target cell, respectively, and $Q_{\text{ratio}}$ is the ratio of accumulated charge for the full and empty target data runs. For this dataset, $Q_{\text{ratio}} = 13.67$.

\item Inaccessible kinematic holes $\Delta^7N_{\text{EH}}$. In order to extract the integrated single-differential cross sections, see Section~\ref{sec:extraction}, the fully differential (7D) cross sections and, hence, the experimental yields in all possible 7D bins, must be obtained. But not all of these 7D kinematic bins can be directly filled using the experimental data because the kinematics of part of these bins are correlated with physical holes in the detector. The yield in these kinematical holes, $\Delta^7N_{\text{EH}}$ from Eq.~\ref{eqn7Dxsec}, has to be estimated using the simulation, and was obtained from the hole yields determined in the simulation. The underlying idea of this process is that while the thrown (or generated) $ep \to e'\pi^+\pi^-p'$ events cover the maximum allowed 7D binned phase space, due to the physical holes in the detector that are modeled in the simulation, the events in some of these 7D bins will not register in the detector. Therefore, some of the 7D bins will be empty and are called kinematical holes. Holes are defined by the acceptance, see Eq.~\ref{eqnSA7}, being zero. The yields in these kinematical holes were obtained from the thrown MC event yields and were appropriately scaled to fill the experimental data. The scale factor was defined by the ratio of the acceptance-corrected reconstructed over the thrown simulated yields, both integrated over all 7D bins that are not kinematical holes. The thrown simulated yield in each 7D hole was then scaled by this ratio and used as the new experimental yield in the 7D hole. This process is called hole filling. Further details can be found in Ref.~\cite{ananote}.

\end{enumerate}

The remaining terms to determine the cross sections from the binned yields include the virtual photon flux factor (see Eq.~\ref{eq:flux}) and the beam-target integrated luminosity. The integrated luminosity used to normalize the total number of acceptance corrected events in each 7D bin to obtain the 7D differential cross sections was determined using 

\begin{equation}
\label{eqnLuminosity}
	{\cal L}=\frac{Q_{\text{tot}}l_t \rho_t N_A}{q_e M_H},
\end{equation}
\noindent
where $Q_{\text{tot}} = 21.32$~mC is the total incident charge on the target, $l_t$ is the length of the target (5.0~cm), $\rho_t$ is the density of liquid hydrogen (0.073~gm/cm$^3$), $N_A$ is Avogadro's number, $q_e$ is the elementary charge, and $M_H$ is the molar mass of hydrogen. For the dataset used for this analysis the total integrated luminosity was 28.18~fb$^{-1}$.

\subsection{Data Binning}
\label{binning}

The 5D cross section defined in Eq.~\ref{eqn5Dxsec} was determined in each bin of $Q^2$ and $W$ defined for the data analysis. The data was sorted into five $Q^2$ bins

\begin{equation*}
[2.0,2.4], [2.4,3.0], [3.0,3.5], [3.5,4.2], [4.2,5.0]~{\rm GeV}^2.
\end{equation*}
The $W$ range of the data spanned from 1.400 to 2.125~GeV and was sorted with a bin size of $\Delta W$=25~MeV. The remaining $\Delta^5\tau$ phase space element in the hadronic kinematic variable was given by

\begin{equation}
\Delta \tau_{\pi^-} = \Delta M_{p'\pi^+} \Delta M_{\pi^+\pi^-} \Delta \cos \theta_{\pi^-} \Delta \phi_{\pi^-} \Delta \alpha_{[\pi^-p][\pi^+p']}.
\end{equation}
\noindent
In the data analysis, the nominal 5D bin sizes were given by

\begin{itemize}
\item $\Delta M_{p'\pi^+}$ = 25~MeV (14 bins)
\item $\Delta M_{\pi^+\pi^-}$ = 25~MeV (14 bins)
\item $\Delta \theta_{\pi^-}$ = 18$^\circ$ (10 bins)
\item $\Delta \phi_{\pi^-}$ = 18$^\circ$ (10 bins)
\item $\Delta \alpha_{[\pi^-p][\pi^+p']}$ = 18$^\circ$ (10 bins).
\end{itemize}

Given the dimensionality of the 5D phase space and the chosen bin sizes, each $(Q^2,W)$ bin was divided into 196,000 cells. The final set of one-fold differential cross sections for the final analysis was obtained by integration of the 5-fold differential cross sections over the four hadronic variables in each $Q^2$ and $W$ bin. Full details on the integral definitions are provided in Ref.~\cite{Isupov:2017lnd} (see Eqs. 17 and 18).

\subsection{Acceptance}
\label{sec:acc}

All the cuts applied were necessary not just for selecting the \fullrctn{} \allowbreak events from all possible events, but also to use only the fully functional areas of the detector. In this process some events were lost and must be accounted for in order to obtain the cross sections. The acceptance correction factor that accounts for these lost events was obtained from the simulation process in a model-independent way, where the \fullrctn{} reaction was simulated in accordance with the experimental conditions. Cuts defined to isolate the \fullrctn{} \allowbreak reaction in the experimental data were also defined for the simulated data and then applied to the simulated events. Since the number of simulated events was known, the fraction of events lost due to the cuts could be obtained. The quantitative values of the calculated acceptance and related quantities from which it is calculated are described in Section~\ref{sctn_acceptance_calculation}. Essential advances in the acceptance extraction for the $\pi^+\pi^-p$ final state that unambiguously establish, for the first time, the required level of simulation statistics needed to extract stable cross sections and to significantly reduce the remaining systematic acceptance uncertainty associated with the simulation statistics are presented in Sections \ref{sctn_rqrd_simstats} and \ref{sctn_cut_rel_err_SA}.

\subsubsection{Acceptance Calculation} 
\label{sctn_acceptance_calculation}

The acceptance was obtained from MC in each of the 7D bins of the analysis as the ratio of the total number of reconstructed events that survive all cuts to the total number of generated events as
\begin{equation} 
\label{eqnSA7}
  SA^7=\frac{SR^7}{ST^7}.
\end{equation}

\noindent
Here $ST^7$, $SR^7$, and $SA^7$  are total the number of thrown events, the total number of reconstructed events that survive all cuts, and the acceptance in each 7D bin, respectively. (The letters `S', `T', `R', and `A' in this nomenclature are used to denote simulation, thrown, reconstructed, and acceptance, respectively.)

\begin{figure}[htb] 
    \includegraphics[width=0.5\textwidth]{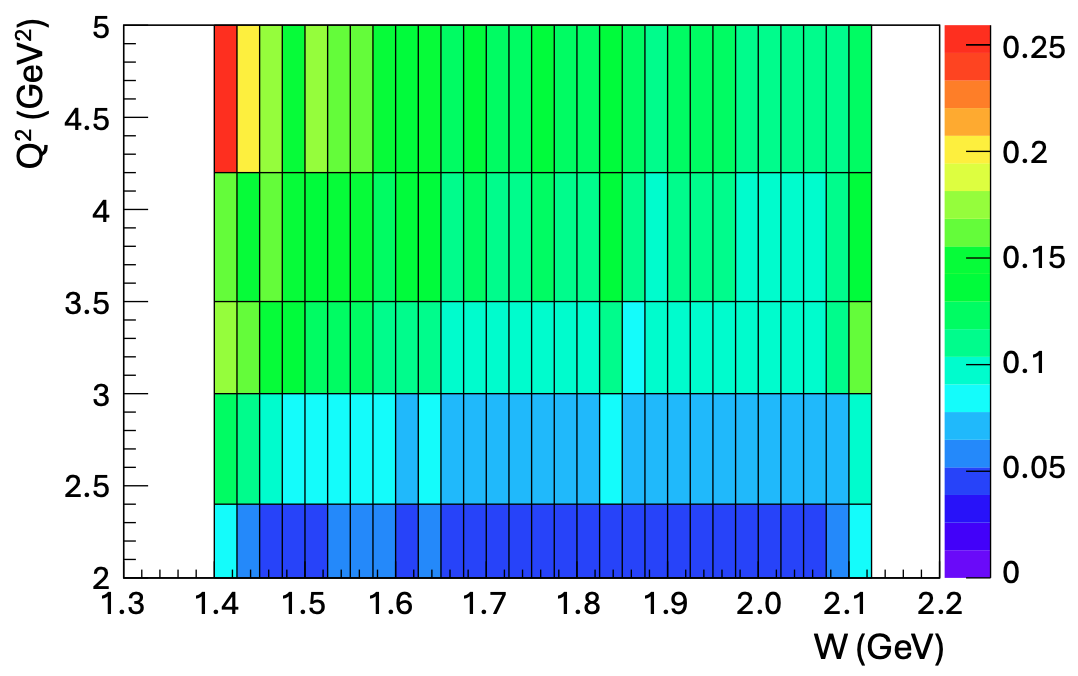}
    \caption{Average 5D acceptance for \fullrctn{} events within each $(Q^2,W)$ of this analysis computed using only the 5D bins that are filled in the experimental phase space.}
    \label{fig_avSA_ER_PS}
\end{figure}

Figure~\ref{fig_avSA_ER_PS} shows the average 5D acceptance per $(Q^2,W)$, labeled $\langle SA^5 \rangle$, computed in only those bins that are filled in the experimental data (labeled $\langle ER^5 \rangle$) using Eq.~\ref{eqnSA7}. (The corresponding figures for $\langle ST^5 \rangle$ and $\langle SR^5 \rangle$ can be found in Ref.~\cite{ananote}.) It is important to note that only the 5D acceptance in the experimentally filled phase space was required to estimate the true experimental yield. The simulation, however, was carried out over the full theoretically accessible kinematic phase space because this data was needed to obtain the hole-filled cross sections as discussed in Section~\ref{corr-yields}.3.

\subsubsection{Required Level of Simulation Statistics} 
\label{sctn_rqrd_simstats}

In comparison to the two-particle final state of the single-meson electroproduction channel, the three-particle final state of the double-pion electroproduction channel has a significantly larger and more complex phase space. Furthermore, electroproduction measurements have many more kinematic bins compared to photoproduction. Both of these aspects, which lead to orders of magnitude higher simulation statistics demand, were systematically studied and found to affect, in a way neither understood nor noted before, the significantly higher level of simulation statistics needed for the accurate extraction of the acceptance values.

\begin{figure*}[htb]
\centering
\includegraphics[width=0.7\textwidth]{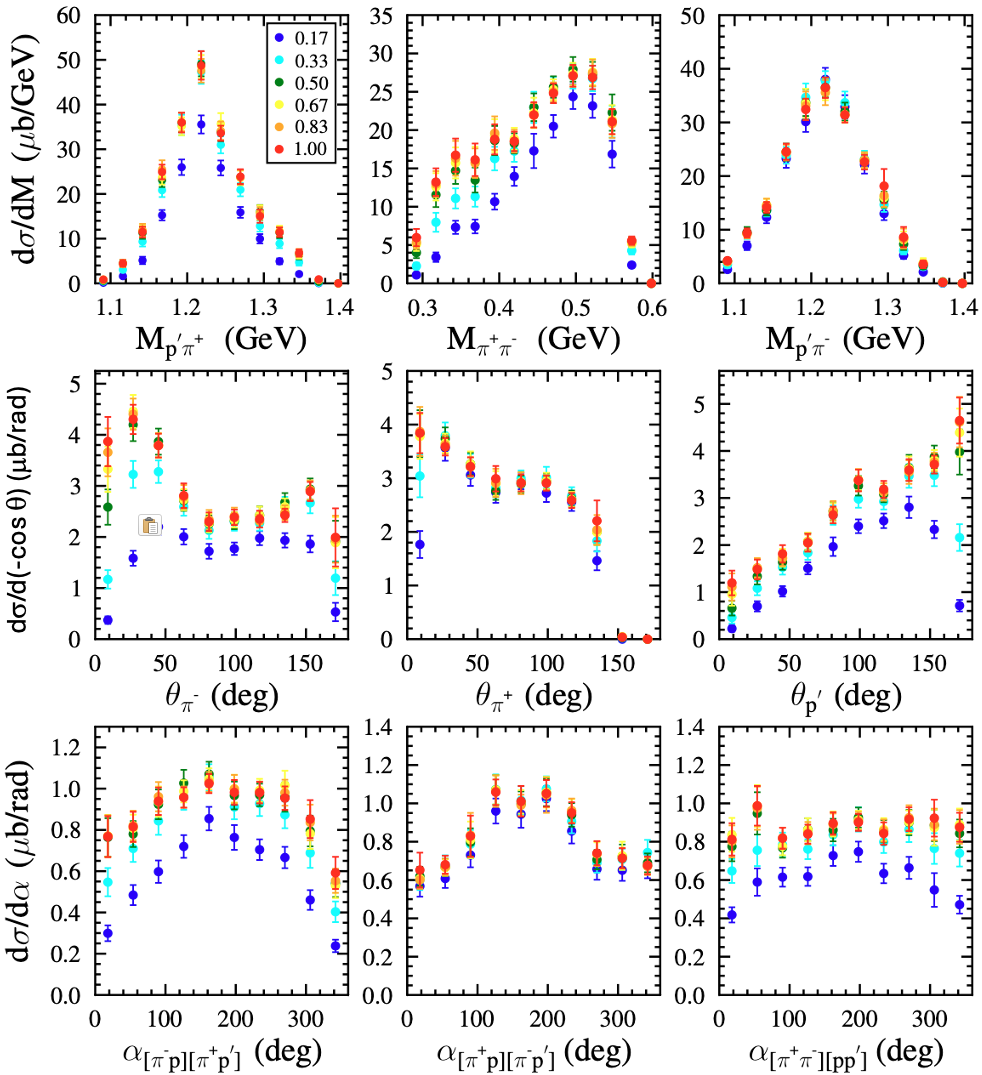}
\caption{Single-differential cross sections {\em without} hole-filling for $Q^2 = [2.00,2.40]$~GeV$^2$ and $W = [1.500,1.525]$~GeV as a function of the fraction of the total simulation statistics used in this analysis from 17\% to the full 100\% sample used for the final cross section analysis. See the MC sample fraction key on the plot.}
\label{fig_obs1D_EC_fncn_simstats}
\end{figure*}

\begin{figure*}[htb]
\centering
\includegraphics[width=0.7\textwidth]{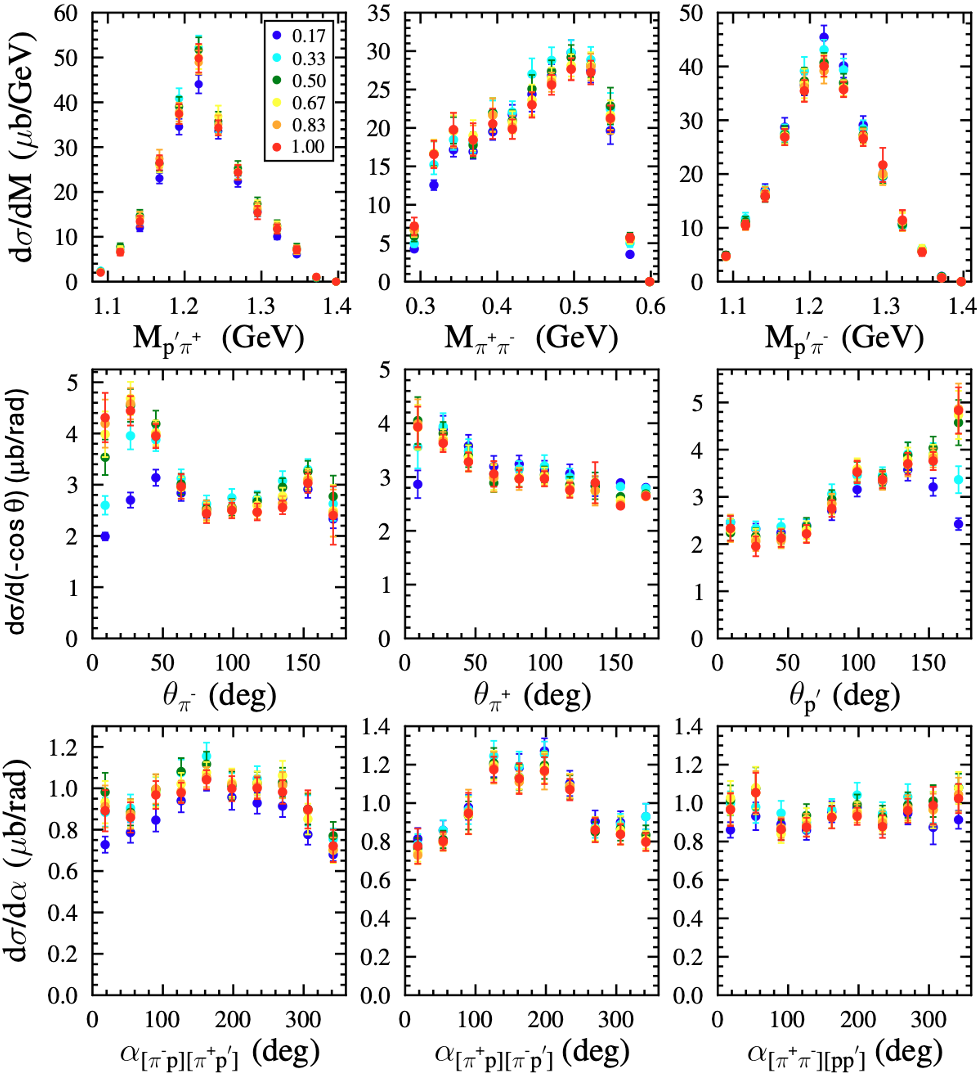}
\caption{Single-differential cross sections {\em with} hole-filling for $Q^2 =[2.00,2.40]$~GeV$^2$ and $W = [1.500,1.525]$~GeV as a function of the fraction of the total simulation statistics used in this analysis from 17\% to the full 100\% sample used for the final cross section analysis. See the MC sample fraction key on the plot.}
\label{fig_obs1D_EF_fncn_simstats}
\end{figure*}

As the simulation statistics reaches the required level, the cross sections and their related uncertainties become stable and no longer change with a further increase in statistics. This can be seen in Figs.~\ref{fig_obs1D_EC_fncn_simstats} and \ref{fig_obs1D_EF_fncn_simstats} for the single-differential cross sections for $Q^2 = [2.40,3.00]$~GeV$^2$ and $W = [ 1.725,1.750]$~GeV without and with hole-filling, respectively, where these cross sections are plotted as a function of the fraction of the total simulation statistics used in this analysis. It can be seen that as this fraction increases up to 1, the cross sections and their related statistical uncertainties become stable. 

\begin{figure}[htb] 
\includegraphics[width=0.45\textwidth]{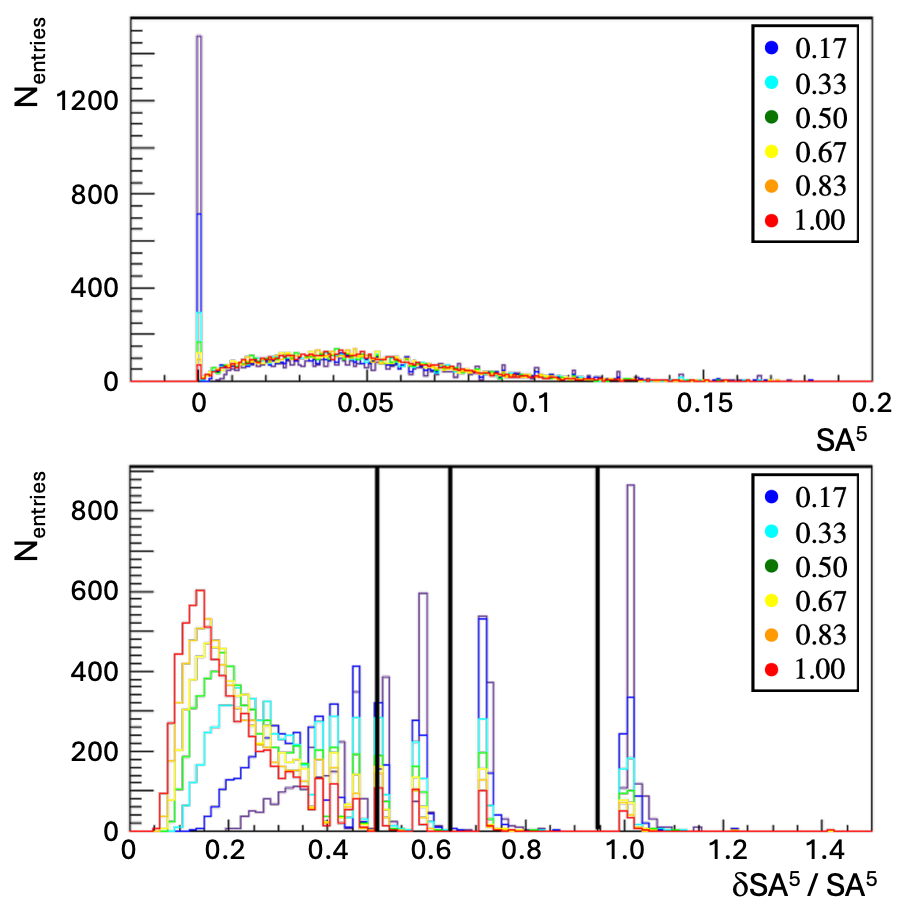}
\caption{Distribution of the acceptance $SA^5$ (top) and the relative uncertainty of $SA^5$ (bottom) for $Q^2 = [2.00,2.40]$~GeV$^2$, $W = [1.650,1.675]$~GeV. The three black lines from left to right in the bottom plot depict the three cut values of 0.50, 0.65, and 0.95, respectively, analyzed for the cut on the relative uncertainty of $SA^5$. The color coding in this figure is the same as in Figs.~\ref{fig_obs1D_EC_fncn_simstats} and \ref{fig_obs1D_EF_fncn_simstats} to represent different fractions of the full simulation statistics included in the plots.}
\label{fig_rel-err-SA}
\end{figure}

\subsubsection{Relative Acceptance Uncertainty Cut} 
\label{sctn_cut_rel_err_SA}

Figure~\ref{fig_rel-err-SA} shows the acceptance $SA^5$ and its relative uncertainty for a representative $(Q^2,W)$ bin. These plots reveal two prominent features. The distribution of the acceptance over all of the 5D bins has a peak at zero. Also, the distribution of the relative uncertainty shows several peaks above 0.5. Both features result from simulation statistics in the corresponding 5D bins that are too small. They cause an unreliable determination of the acceptance and, hence, the cross section. However, they become less pronounced as the simulation statistics increase. 

Even when the simulation statistics approaches the required level to provide stable measured cross sections, a further reduction of the impact of the still remaining 5D phase space bins with very few events can be achieved by cutting on the relative acceptance uncertainty and by treating the removed 5D bins as kinematic holes. After a relative acceptance uncertainty cut is applied to the red distribution (representing the highest statistics) in the bottom half of Fig.~\ref{fig_rel-err-SA}, the extracted fully integrated and one-fold differential cross sections continue to be consistent within the quoted error bars but become even less sensitive to the simulation statistics. When optimizing the relative acceptance uncertainty cut, the impact of a higher systematic uncertainty due to hole-filling with the large statistical fluctuations due to large relative acceptance uncertainties must be balanced. The three black lines from left to right in the bottom half of Fig.~\ref{fig_rel-err-SA} depict the three studied cut values of 0.50, 0.65, and 0.95, respectively, and the cut value of 0.65 that best meets this balance~\cite{ananote}.

\subsection{Systematic Uncertainties}

The individual sources of systematic uncertainty in this analysis fall within several broad categories, those that can be i) directly estimated or are ii) simulation driven and those that depend on iii) scale effects. The first two categories were assigned as point-to-point systematic uncertainties and the final category represents overall scale uncertainties. In this section a brief overview of the individual sources is provided along with insight into how the uncertainties were estimated. Table~\ref{table_SE} provides a complete summary. 

Since most of the estimated systematic uncertainties of the cross sections were determined by varying the extraction conditions, they were dominated by and, hence, well contained within their statistical uncertainties, which were dominantly driven by the experimental yields. The remaining estimated systematic uncertainties were furthermore mostly uncorrelated. The total systematic uncertainty for the analysis is therefore obtained through standard Gaussian error propagation,
\begin{equation} 
\label{eqnGaussErrProp}
  \sigma_{\text{tot}}=\sqrt{\sum_{i=1} {\sigma_{i}}^{2}},
\end{equation}
\noindent
where $\sigma_{\text{tot}}$ is the total estimated systematic uncertainty of this analysis and $\sigma_{i}$ are the individual $(Q^2,W)$ averaged systematic uncertainties listed in Table \ref{table_SE}. The estimated total $(Q^2,W)$ averaged systematic uncertainty summed correspondingly over all individual contributions in quadrature is 11.5\%.

\vskip 0.5cm
\noindent
i) \underline{Directly Estimated}:
The systematic uncertainties for sources \#1 to \#7 in Table \ref{table_SE} were first analyzed in each $(Q^2,W)$ and the listed uncertainties were then obtained by averaging over all $(Q^2,W)$ bins. They are dominated by and contained well within their statistical uncertainties. A statistically driven process was used where the individual systematic uncertainty was calculated as the standard deviation of the variation of the cross section measurement under variation of the extraction cuts within realistic limits (\#1 to \#6) or in dependence on the selected independent-variable set (\#7). To mitigate the statistical impact on the estimated systematic uncertainties, only integrated single-differential cross sections (Section \ref{sec:obs_1D}) and photon-polarization-dependent observables (Section \ref{sec:obs_R2}) were used at the level of each $(Q^2,W)$, as the integrated yields would be otherwise distributed over 1D and 2D binned phase spaces. 

The effect of the variable set dependent extraction of cross sections, item \#7 in the Table \ref{table_SE}, can only be considered after the cross sections have been extracted using all core tasks. This effect is observed by noting that the integrated \vprctn cross sections for a $(Q^2,W)$, obtained from each of the three independent-variable sets (Section \ref{sec:formalism}), are not the same.

\vskip 0.5cm
\noindent
ii) \underline{Simulation Driven}:
Effects of some systematic uncertainties on the cross sections are based on the simulation and only observed indirectly, as in the cases of hole filling, item \#8 in Table \ref{table_SE}, and background subtraction and $MM^2$ distribution differences between measured and simulated data, items \#9 and \#10 in Table \ref{table_SE}, respectively. 

The systematic uncertainty for the background contribution due to the leakage of $3\pi$ events into the nominal $MM^2(\pi^-X)$ cut range was determined by the fitting and extrapolation within the low $W$ uncertainties of the normalization scales of the simulated $2\pi$ and $3\pi$ contributions that described the total measured $MM^2(\pi^-X)$ yields best. However, as the contribution was already very small, the average systematic was assigned as 1\%. The systematic uncertainty due to adapting the simulated to the measured resolution and due to subsequent $MM^2(\pi^-X)$ event selection cuts was assigned as 2\% by varying the smearing parameters and the cut width within realistic limits. To maximize statistics and reduce the statistical uncertainty dominance on the estimated systematic uncertainties, both were calculated at the level of the single-differential integrated data in each $(Q^2,W)$.

For hole filling, $\Delta^7N_{\text{EH}}$ from Eq.~\ref{eqn7Dxsec}, the systematic uncertainty in this analysis can be calculated in each single-differential and photon-polarization-dependent cross section bin. Since the hole filling was based on the two-pion event generator, half of the absolute difference between the cross sections with and without hole-filling was conservatively estimated as the systematic uncertainty. The uncertainty listed under item \#8 in Table \ref{table_SE} is averaged over all 7D bins and as such is statistically the most precisely estimated systematic uncertainty.

\vskip 0.5cm
\noindent
iii) \underline{Scale Dependent}: There are two overall scale uncertainty sources, items \#11 and \#12 of Table \ref{table_SE}. The first is associated with the radiative correction. For this analysis, as for previous analyses, a conservative uncertainty of 5\% was assigned to the Mo and Tsai correction approach applied to the highly-integrated double-pion electroproduction cross sections, see e.g. Refs.~\cite{Isupov:2017lnd,Fedotov:2018oan}. The second is associated with the beam-target luminosity. The associated systematic uncertainty has contributions from the collected beam charge uncertainty, the uncertainty in the target density in terms of target pressure and temperature variations over the run range, and the uncertainty in the length of the cryotarget cell. The assigned systematic uncertainty was set as 5\%.

\begin{table*}[htb]
\centering 
\begin{tabular}{c c c} \hline
Item & Source                                        & Uncertainty \\ \hline
1 & Electron Identification                          & ${\lesssim 1\%}$ \\ 
2 & Electron Fiducial Cut                            & ${\lesssim 1\%}$ \\
3 & Hadron Identification                            & 3.1\% \\
4 & Hadron Fiducial Cut                              & ${\lesssim 1\%}$ \\
5 & Detector Inefficiency Identification             & ${\lesssim 1\%}$ \\ 
6 & Acceptance Calculation                           & ${\lesssim 1\%}$ \\
7 & Variable Set Dependent Extraction                & 5.5\% \\
8 & Hole Filling                                     & 5.7\% \\
9 & Background Subtraction                           & ${\lesssim 1\% }$ \\
10 & Missing Mass Squared Cut                        & 2\% \\
11 & Radiative Effects Correction                    & 5\% \\
12 & Luminosity Measurement                          & 5\% \\ 
\hline
\end{tabular}
\caption{Systematic uncertainties for the extracted cross sections for the exclusive $ep \to e'\pi^+\pi^-p'$ reaction averaged over all $(Q^2,W)$ bins.}
\label{table_SE}
\end{table*}
\section{Results and Discussion}
\label{sec:results}

\subsection{Cross Section Results}

\begin{figure}[htb]
  \centering
  \includegraphics[width=0.95\columnwidth]{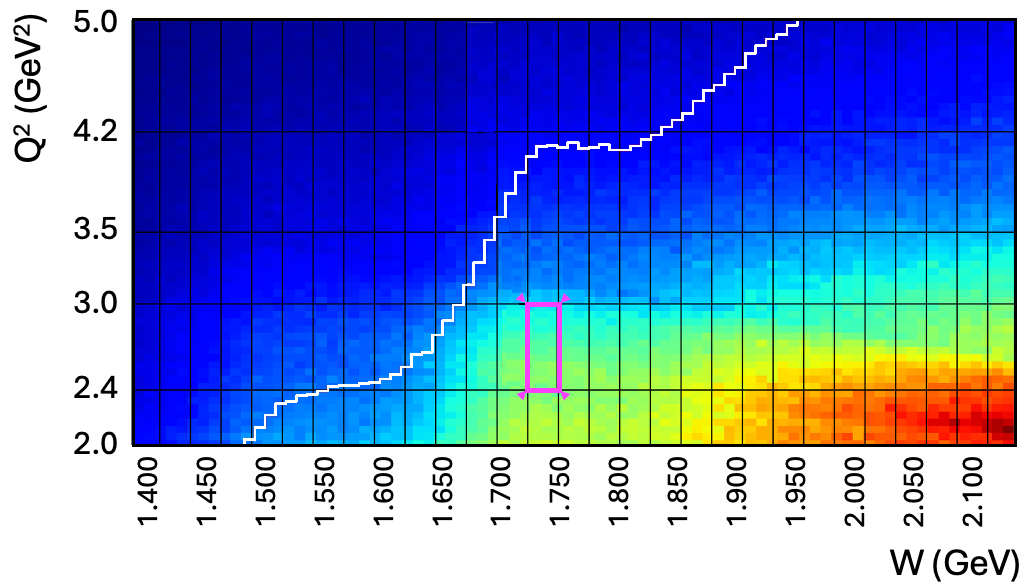}
  \caption{Yield distribution in terms of $Q^2$ vs. $W$ for the charged two-pion events in the kinematic region of this analysis with the kinematic binning grid overlaid. The representative bin ($Q^2 = [2.40,3.00]$~GeV$^2$, $W = [1.725,1.750]$~GeV) chosen to display the analysis results is highlighted by the magenta box. The white curve running through the data shows a 1D projection of the $W$ distribution highlighting the $N^*$ resonance enhancements.}
  \label{fig_q2w_coverage}
\end{figure}

In this section the bin-averaged cross sections that were extracted from this analysis are highlighted. Only the results for a representative bin, $Q^2 = [2.40,3.00]$~GeV$^2$, $W = [1.725,1.750]$~GeV, are shown. The extracted cross sections over the full kinematic phase space are included in the CLAS Physics Database~\cite{clasphysdb}. The results are shown separately for the nine one-fold differential cross sections and the polarization-dependent cross sections. Figure~\ref{fig_q2w_coverage} shows the $Q^2$ vs. $W$ kinematic region for this analysis with the nominal binning grid overlaid.

\begin{figure*}[htb]
  \centering
  \includegraphics[width=0.7\textwidth]{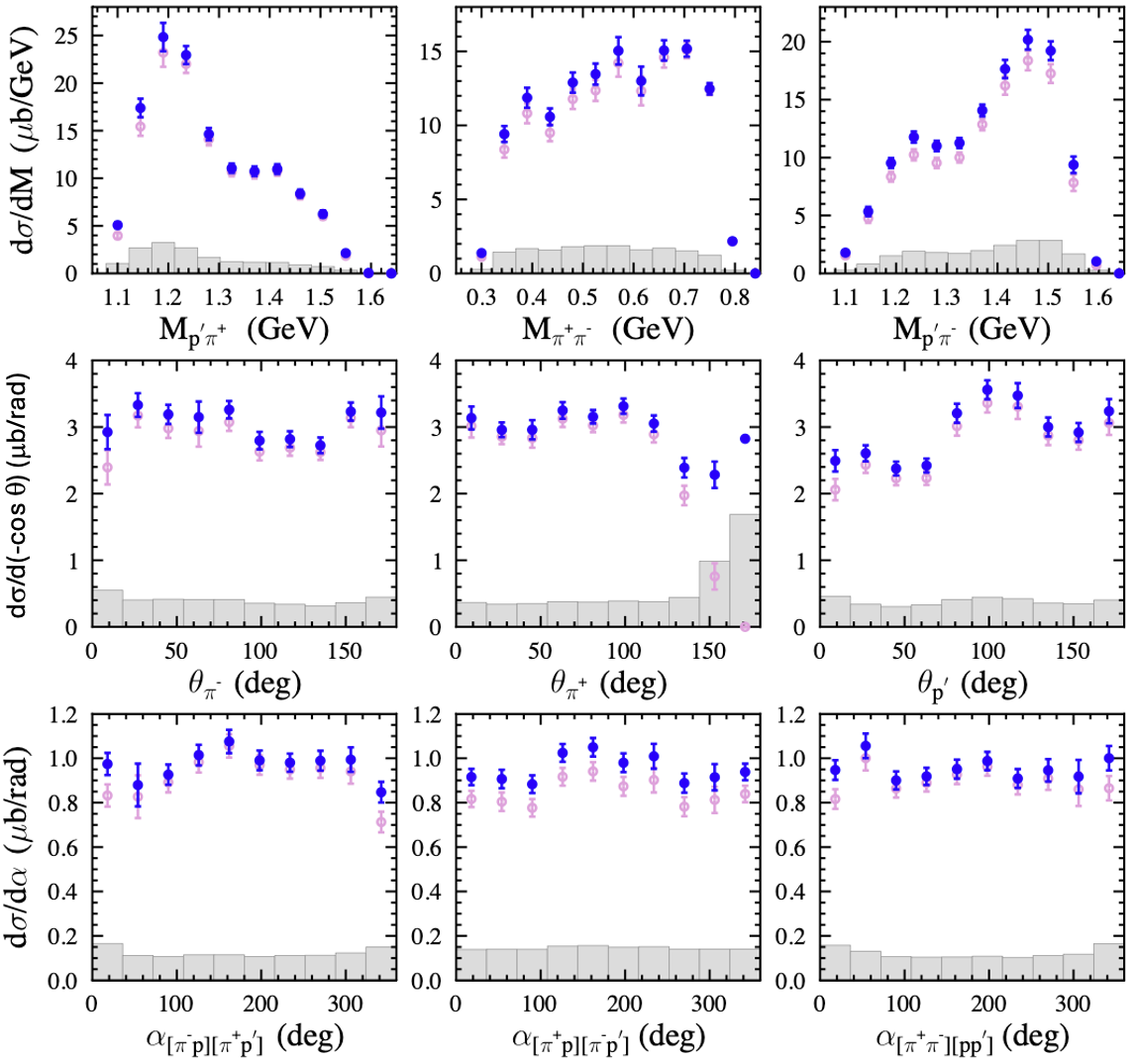}
  \caption{Single differential cross sections for the representative $(Q^2,W)$ bin $Q^2 = [2.40,3.00]$~GeV$^2$, $W = [1.725,1.750]$~GeV. The light magenta open circles and the blue solid circles show the experimentally measured cross sections without and with the kinematical hole filling, respectively. The shaded histogram at the bottom of each subplot shows the assigned systematic uncertainty.}
  \label{fig_result_1D}
\end{figure*}

Figure~\ref{fig_result_1D} shows the nine one-fold differential cross sections defined in Section~\ref{sec:obs_1D} for the representative $(Q^2,W)$ bin. The light magenta open circles and solid blue points show the measured cross sections without and with the kinematical holes filled, respectively, using the MC simulation as discussed in Section~\ref{sec:analysis}. The corresponding colored error bars show the statistical uncertainties and the black shaded histogram at the bottom of each subplot shows the systematic uncertainties.

\begin{figure*}[htb]
  \centering
  \includegraphics[width=0.7\textwidth]{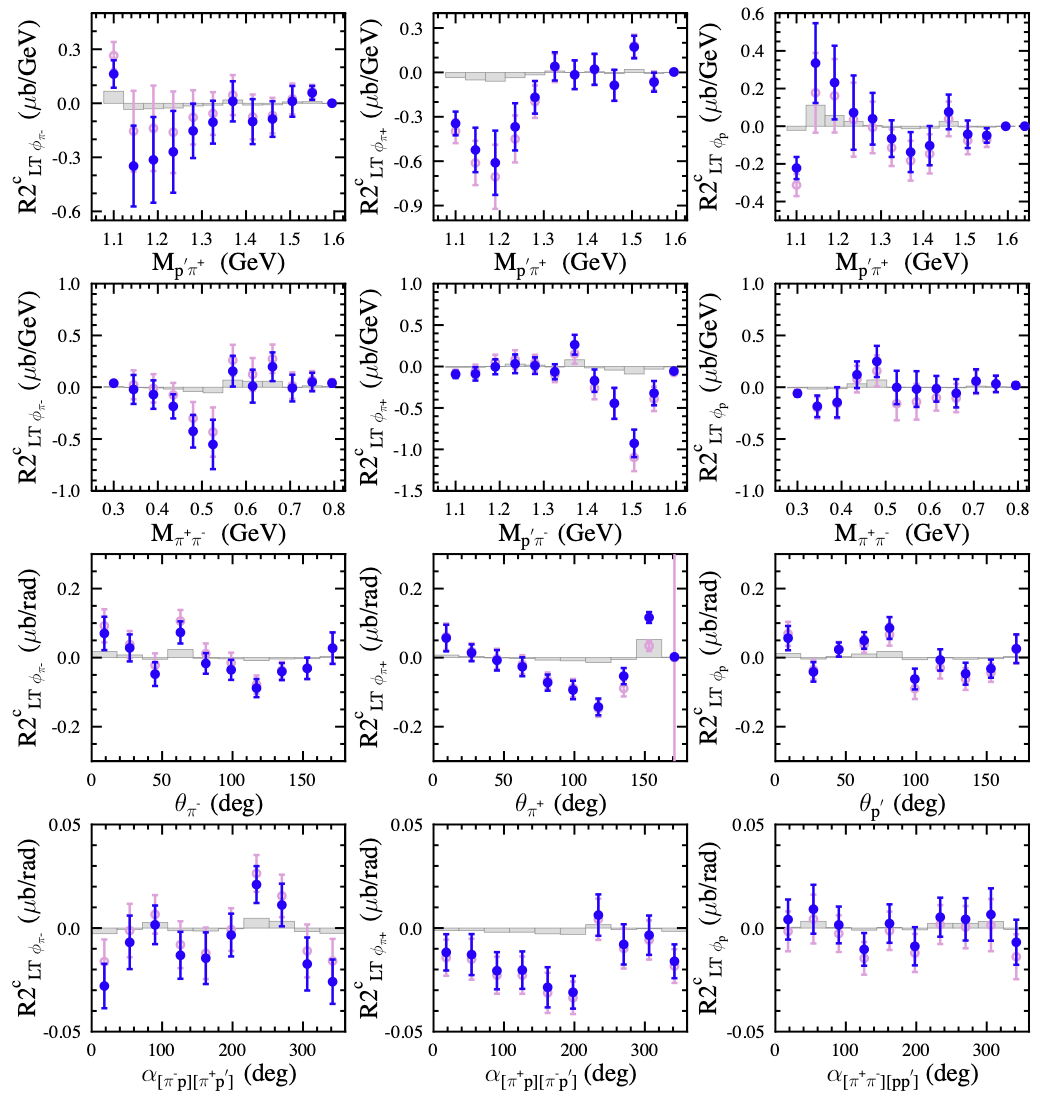}
  \caption{\RB{} for the representative $(Q^2,W)$ bin $Q^2 = [2.40,3.00]$~GeV$^2$, $W = [1.725,1.750]$~GeV. The light magenta open circles and the blue solid circles show the experimentally measured cross sections without and with the kinematical hole filling, respectively. The shaded histogram within each subplot shows the assigned systematic uncertainty.}
  \label{fig_result_R2B}
\end{figure*}

\begin{figure*}[htb]
  \centering
  \includegraphics[width=0.7\textwidth]{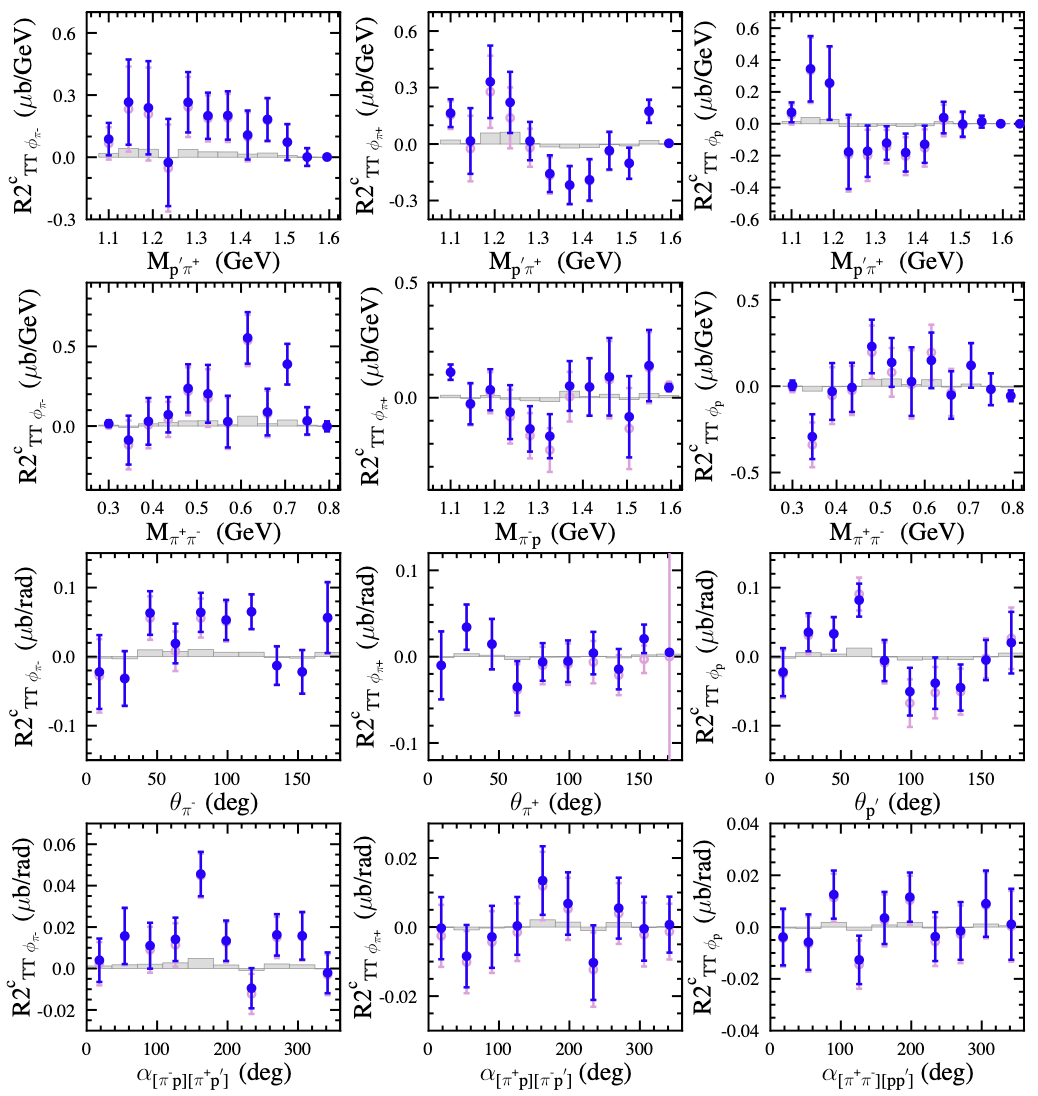}
  \caption{\RC{} for the representative $(Q^2,W)$ bin $Q^2 = [2.40,3.00]$~GeV$^2$, $W = [1.725,1.750]$~GeV. The light magenta open circles and the blue solid circles show the experimentally measured cross sections without and with the kinematical hole filling, respectively. The shaded histogram within each subplot shows the assigned systematic uncertainty.}
  \label{fig_result_R2C}
\end{figure*}

\begin{figure*}[htb]
  \centering
  \includegraphics[width=0.7\textwidth]{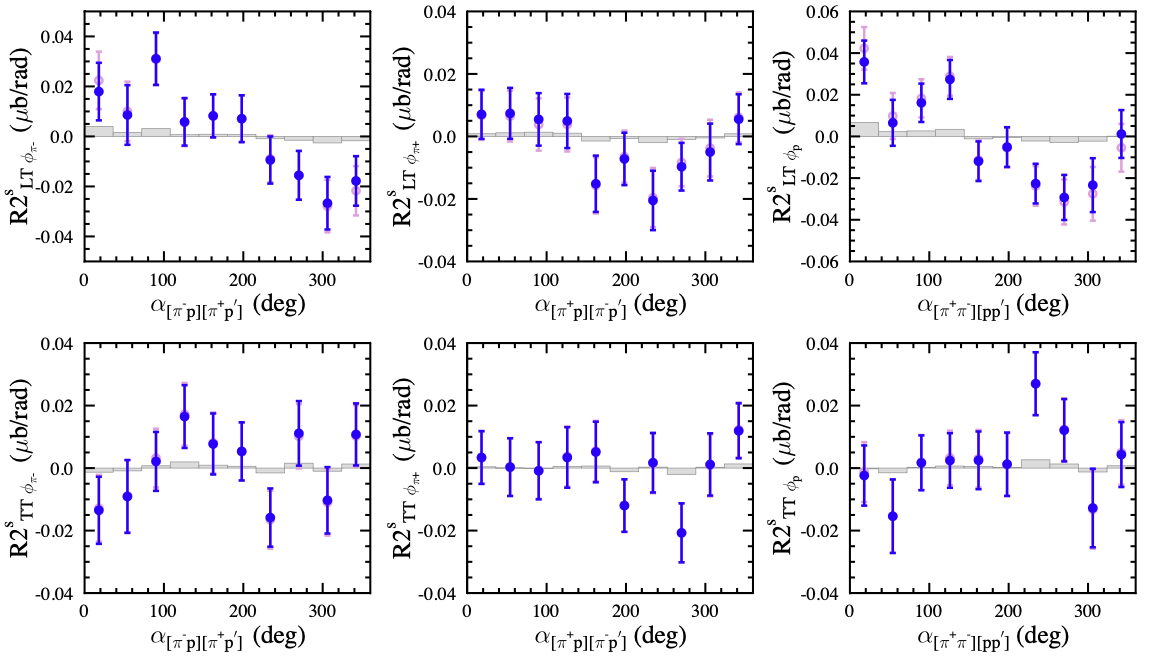}
  \caption{\RD{} (top) and \RE{} (bottom) for the representative $(Q^2,W)$ bin $Q^2 = [2.40,3.00]$~GeV$^2$, $W = [1.725,1.750]$~GeV. The light magenta open circles and the blue solid circles show the experimentally measured cross sections without and with the kinematical hole filling, respectively. The shaded histogram within each subplot shows the assigned systematic uncertainty.}
  \label{fig_result_R2DE}
\end{figure*}

Figures~\ref{fig_result_R2B}-\ref{fig_result_R2DE} show the polarization-dependent cross sections defined in Section~\ref{sec:obs_R2} for the representative $(Q^2,W)$ bin similarly as for Fig.~\ref{fig_result_1D}. These figures show the cross  sections denoted as \RB{}, \RC{}, \RD{}, and \RE{}. Note that a separate plot for \RA{} is not shown as it is equivalent to Fig.~\ref{fig_result_1D} (apart from a factor of $2\pi$).

\begin{figure*}[htb]
  \centering
  \includegraphics[width=0.7\textwidth]{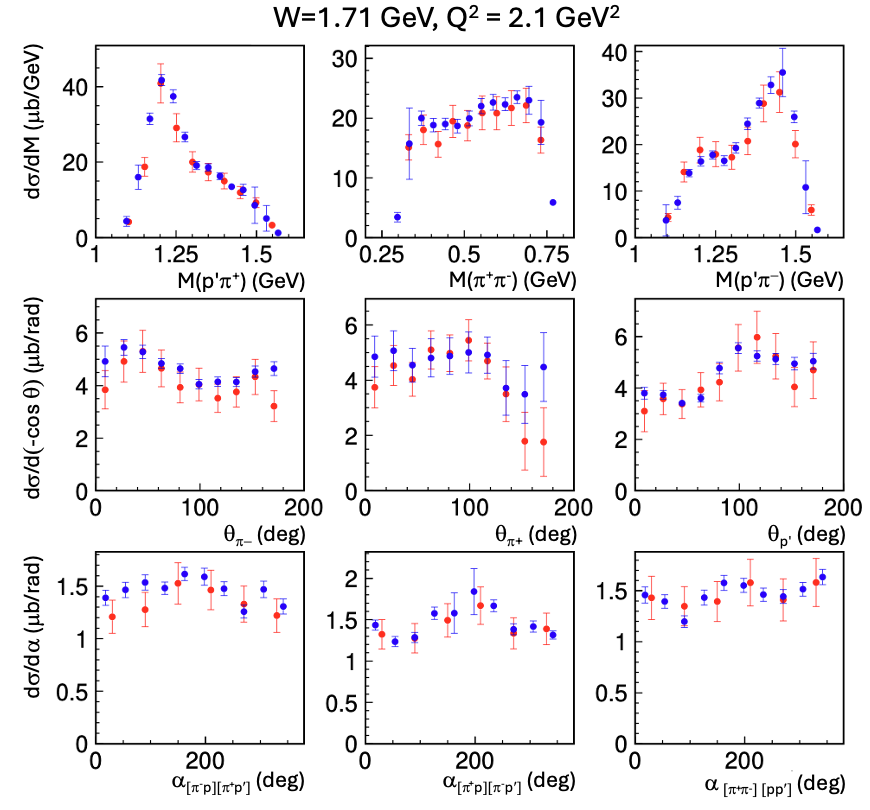}
  \caption{Single-differential cross sections for a representative bin $Q^2 = [2.00,2.40]$~GeV$^2$, $W = [1.725,1.750]$~GeV comparing the previous CLAS results from Ref.~\cite{Isupov:2017lnd} (red) to those of this analysis (blue). Only the statistical uncertainties are shown for both datasets.}
  \label{arjun_isupov}
\end{figure*}

\subsection{$\pi^+\pi^-p$ Electroproduction Mechanisms}
\label{phenom}

These new CLAS data on the nine one-fold differential $\pi^+\pi^-p$ electroproduction cross sections have been obtained with substantially improved accuracy and reliability relative to the available results for this same dataset in Ref.~\cite{Isupov:2017lnd}. For $W > 1.60$~GeV, both datasets are consistent within error bars. As a representative example, Fig.~\ref{arjun_isupov} shows a comparison of the nine one-fold differential cross sections at $Q^2 = [2.0,2.4]$~GeV$^2$ and $W = [1.700,1.725]$~GeV. The differential cross sections from the two different analyses of the same CLAS dataset are in good agreement, even within statistical uncertainties only, except for a few data points in the angular distributions over the $\theta_{\pi^-}$ and $\theta_{\pi^+}$ polar angles, as well as the $\alpha_{[\pi^-p][\pi^+p']}$ angle.

Here and in all other parts of the kinematic coverage, the new results reported in this work have become available with substantial improvements. In particular:

\begin{itemize}
\item[(a)] The MC statistics were increased by approximately an order of magnitude compared to those employed for the data published in Ref.~\cite{Isupov:2017lnd}, allowing us to populate MC events in nearly all 7D cells where measured events are available;

\item[(b)] A new procedure was developed to demonstrate that the achieved MC statistics are sufficient to stabilize the extracted differential cross sections, see Section~\ref {sctn_rqrd_simstats};

\item[(c)] The 7D cells with low efficiency were treated as inefficient areas, see Section~\ref{sctn_cut_rel_err_SA};

\item[(d)] The differential cross sections were extracted with improved accuracy and are available with finer binning over the invariant masses and $\alpha$ angular distributions.
\end{itemize}

Owing to these improvements, we consider the data presented in the current paper as superseding those from the previous analysis reported in Ref.~\cite{Isupov:2017lnd}. 

This new data offer substantial improvements that are important for the extraction of the $\gamma_vpN^*$ electrocouplings and cover the kinematic domain in which dressed quarks become the dominant contributors to hadron structure, enabling an extension of our knowledge of strong-interaction dynamics over the distance scale where the transition from the strongly coupled to perturbative QCD regimes is expected to occur \cite{Burkert:2025coj,Achenbach:2025kfx,Mokeev:2015lda,Mokeev:2023zhq,Burkert:2019bhp}.

As a first step toward these objectives, we examine the capability of the JLab-Moscow State University JM23 reaction model \cite{Mokeev:2023zhq} to describe the nine one-fold differential cross sections reported in this work over the kinematic range $Q^2 = 2.0\text{–}5.0~\mathrm{GeV}^2$ and $W = 1.4\text{–}2.0~\mathrm{GeV}$. The non-resonant parameters of the JM23 model were varied independently in each $(Q^2,W)$ bin in the description of the data. Their optimal values were determined by minimizing the fit $\chi^2$, calculated point-by-point from comparisons between the measured and calculated nine one-fold differential cross sections within each $(Q^2,W)$ bin. The $N^*$ parameters for the resonances in the mass range below 1.6~GeV were taken from the CLAS analyses of $\pi N$ and $\pi^+\pi^- p$ electroproduction data summarized in Ref.~\cite{Mokeev:2023zhq} and were kept fixed within their established uncertainties. For the $N^*$ states in the mass range above 1.6~GeV, the starting values of their parameters were taken from Ref.~\cite{HillerBlin:2019jgp}. The $\gamma_vpN^*$ electrocouplings were further adjusted to describe the nine independent one-fold differential cross sections in each bin of $(Q^2,W)$ covered by the $\pi^+\pi^-p$ electroproduction data.

\begin{figure}[htp]
\begin{center}
\includegraphics[width=1.0\columnwidth]{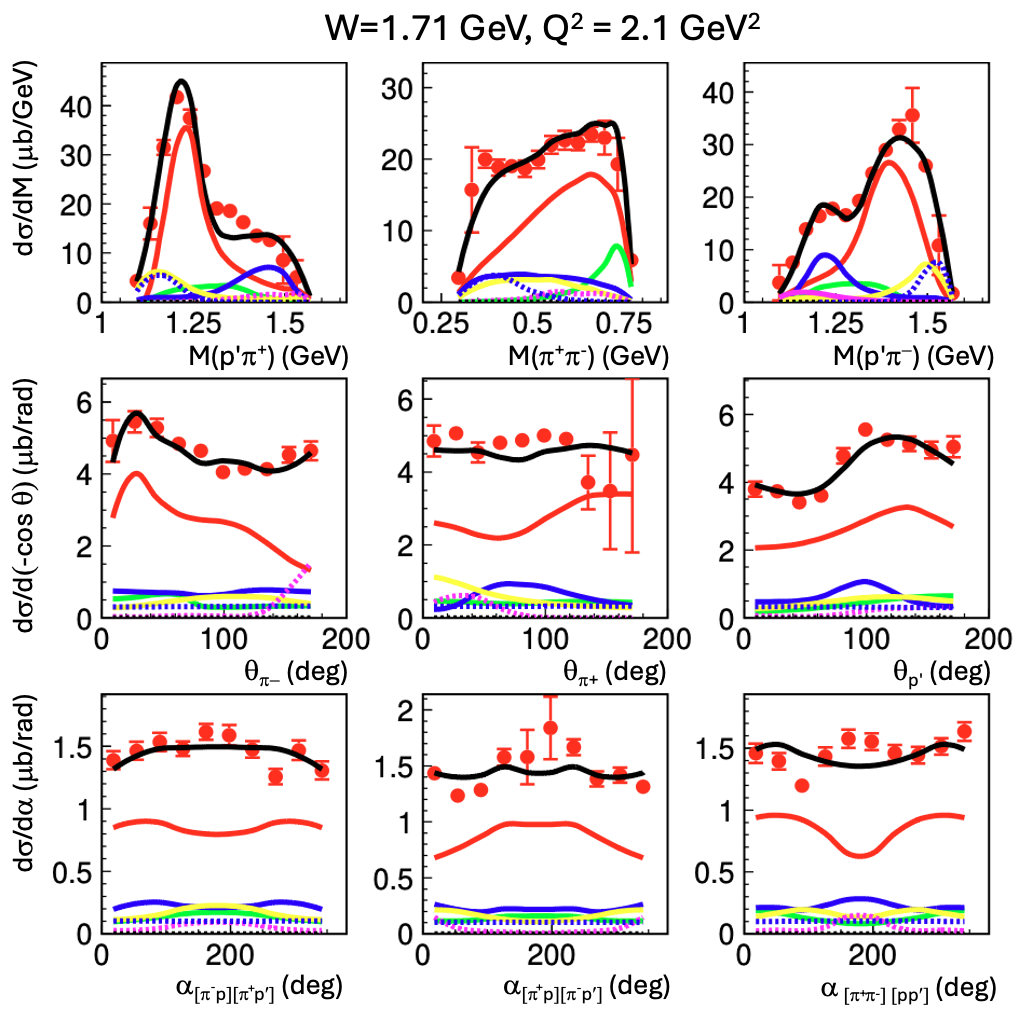}
\caption{Description of the nine one-fold differential cross sections (thick black solid lines) for the $Q^2$-interval from 2.0--2.4 GeV$^2$ for the $W$ bin at 1.70--1.73~GeV within the JM23 model and the associated contributions from the relevant meson-baryon channels: $\pi^-\Delta^{++}$ (red lines), $\rho p$ (green lines), $\pi^+\Delta^0$ (blue lines), $\pi^+N(1520)3/2^-$ (yellow lines), $2\pi$-direct production (dotted magenta lines), and $\pi^+N(1685)5/2^-$ (dotted blue solid lines). For the data uncertainties only statistical contributions are shown.}
\label{channels_21b}
\end{center}
\end{figure}

The $\pi^+\pi^- p$ differential cross sections of these new CLAS data are well described by the mechanisms included in the JM23 model. No additional reaction mechanisms were required to achieve this level of agreement. A representative example of the data description is shown in Fig.~\ref{channels_21b}, where the measured differential cross sections are shown together with those computed from JM23 along with the corresponding contributions from the individual meson-baryon channels. No deficiencies are observed in describing the shapes of the differential cross sections that would manifest systematically over the ranges of $Q^2$ and/or $W$. The individual contributions from the relevant reaction mechanisms are seen to exhibit distinct behaviors in all nine measured one-fold differential cross sections. This enables separation of the amplitudes of the contributing mechanisms and allows for computation of the cross sections associated with each of them directly from their respective amplitudes. The results for the meson-baryon channel contributions available both at the amplitude and the differential/integrated cross section levels may be of interest for testing advanced coupled-channel approaches that account for the intermediate meson-baryon states revealed in $\pi^+\pi^-p$ electroproduction.

\begin{figure}[htp]
\begin{center}
\includegraphics[width=1.0\columnwidth]{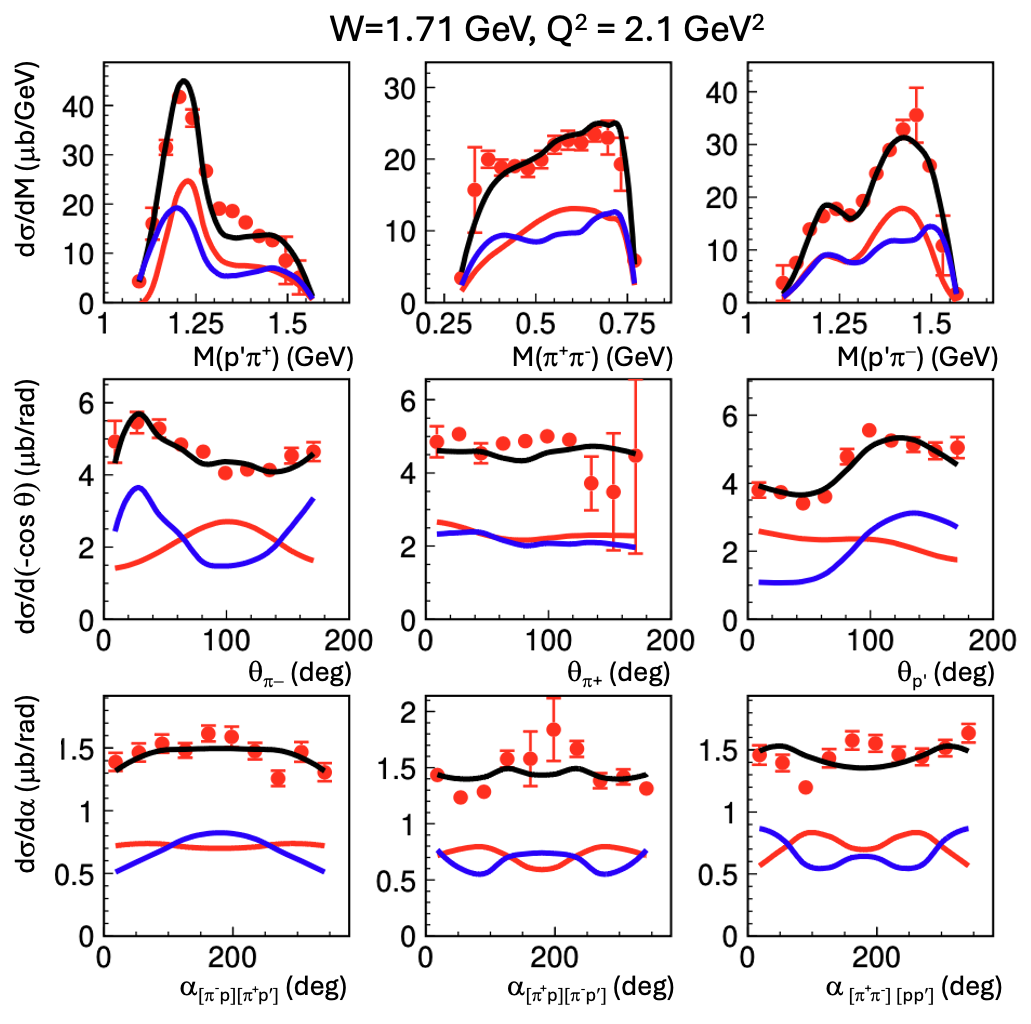}\\
\caption{Resonant (red solid lines) and non-resonant (blue lines) contributions in the nine differential cross sections (black lines) for the $Q^2$-interval from 2.0--2.4 GeV$^2$ for the $W$ bin at 1.70--1.73~GeV. For the data uncertainties only statistical contributions are shown.}
\label{rescontr}
\end{center}
\end{figure}

For each computed differential cross section, the resonant and non-resonant contributions were evaluated by switching off the corresponding amplitudes. The resulting components are shown in Fig.~\ref{rescontr} for a representative $(Q^2,W)$ bin. The magnitudes of the resonant and non-resonant contributions are comparable, but their kinematic dependencies differ noticeably, particularly in the angular distributions. The interference between the resonant and non-resonant amplitudes also has a visible impact on the shapes of the angular distributions. These features provide favorable conditions for separating these contributions, which is essential for extracting the resonance electrocouplings from these data.

\begin{figure}[htp]
\begin{center}
\includegraphics[width=1.0\columnwidth]{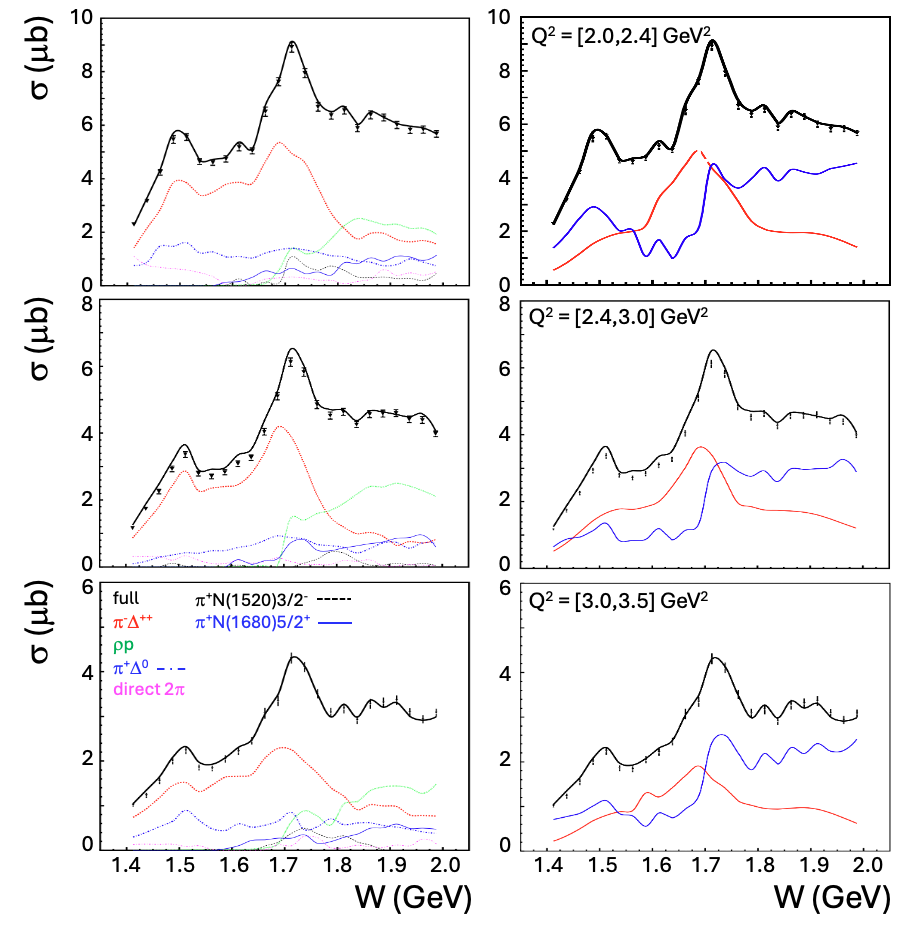}
\caption{Description of the fully integrated $\gamma_v p \to \pi^+\pi^-p'$ cross sections for $Q^2$ from 2.0--3.5~GeV$^2$ and $W$ from 1.4--2.0~GeV. The rows from top to bottom represent different $Q^2$-intervals as labeled. (Left) The contributions from the meson-baryon channels and direct $2\pi$ electroproduction to CLAS data deduced within the JM23 model as labeled. (Right) The separated resonant (red solid) and non-resonant (blue solid) contributions to the cross sections. Only statistical data uncertainties are shown.}
\label{integrated}
\end{center}
\end{figure}

After integrating over the 5D final state hadron kinematic phase space in each $(Q^2,W)$ bin, the fully integrated JM23 cross sections, together with the integrated cross sections for the individual mechanisms, become available. The experimental results for the fully integrated cross sections, together with the predictions of the JM23 model fitted to the data, are shown in the left column of Fig.~\ref{integrated}. The right column of Fig.~\ref{integrated} displays the separated resonant and non-resonant contributions. The parameters of the JM23 model used for the evaluation of the fully integrated cross sections shown here have been adjusted to describe the differential cross section data in the  limited range of $Q^2 < 3.5$~GeV$^2$ and $W < 2.0$~GeV. In a companion paper to the data presented in this work, these new CLAS data have been fit within the JM23 model to extend its range of applicability for $Q^2$ up to 5.0~GeV$^2$~\cite{2pi-hilevel}.

From the measurements reported in this paper, the the interference $TT$ and $LT$ $\pi^+\pi^-p$ electroproduction cross sections have become available for the first time. These components were extracted in terms of the nine one-fold differential cross sections in each $(Q^2,W)$ bin covered by the measurements. This represents a groundbreaking extension of the experimental information, providing direct sensitivity to the detailed interference patterns between the various resonant and non-resonant amplitudes incorporated in the reaction models. Future analyses of these results will substantially deepen our insight into the meson-baryon dynamics underlying $\pi^+\pi^- p$ electroproduction across the entire resonance excitation region. The large body of these observables requires the development of new methods for analysis of the cross section data, in particular, by using artificial intelligence/machine learning approaches.
\section{Conclusions}
\label{sec:conclusions}

Along with 30 new photon-polarization-dependent observables, nine independent one-fold differential and fully integrated $\pi^+\pi^-p$ electroproduction cross sections have been extracted from the measurements with CLAS across the kinematic range $1.40$~GeV $ < W < 2.125$~GeV and $2.0$~GeV$^2 < Q^2 < 5.0$~GeV$^2$. Compared with the previously published CLAS $\pi^+\pi^-p$ electroproduction cross sections extracted from this same dataset in the same kinematic area~\cite{Isupov:2017lnd}, the results presented in this work benefit from a considerably improved and new treatment of the $\pi^+\pi^-p$ event detection efficiency and efficiency uncertainty, respectively. This enabled a more reliable determination of the measured cross sections.

The extracted $\pi^+\pi^-p$ cross sections after implementing this new efficiency uncertainty method remain consistent even when lowering the amount of simulated data down to 50\% of the available simulation statistics. Furthermore, for every populated 7D kinematic cell containing measured $\pi^+\pi^-p$ events, we ensured that a sufficient number of simulated events was available for a reliable determination of the corresponding detection efficiency. Therefore, the $\pi^+\pi^-p$ electroproduction cross sections presented in this paper should be considered as superseding the results from Ref.~\cite{Isupov:2017lnd}.

The data on the virtual photon-polarization-dependent contributions to the nine independent $\pi^+\pi^-p$ electroproduction cross sections -- specifically the $TT$ and $LT$ components of $d\sigma/dM_{i,j}$, $d\sigma/d\cos\theta_i$, and $d\sigma/d\alpha_{[i',j'][i,j]}$ ($i,j = \pi^+, \pi^-, p$) -- have become available for the first time in each $(Q^2,W)$ bin covered by the measurements. These results open important new opportunities for the development of amplitude analyses of $\pi^+\pi^-p$ electroproduction, providing access to detailed interference patterns among the contributing mechanisms. Such interference effects were not extracted from previously available data on this exclusive channel.

A reasonable description of the nine one-fold differential cross sections has been achieved within the JM23 reaction model~\cite{Mokeev:2023zhq} for $2.0 < Q^2 < 5.0$~GeV$^2$ and $1.41 < W < 2.0$~GeV, enabling a separation of the contributions from all meson–baryon channels and direct two-pion electroproduction mechanisms. No evidence has been found for contributions from processes beyond those implemented in the model. 

The separation between the resonant and non-resonant contributions reveals a substantial resonant component across the full range of $W < 2.0$~GeV for the entire $Q^2$ range covered in the measurements. The shapes of the resonant and non-resonant contributions to the nine one-fold differential cross sections are markedly different, especially in the angular distributions. These features provide favorable conditions for the extraction of the $\gamma_v pN^*$ electrocouplings for the contributing $N^*$ states for $2.0 < Q^2 < 5.0$~GeV$^2$ over the mass range $1.6 < W < 2.0$~GeV from this data~\cite{2pi-hilevel}.
\begin{acknowledgments}

The authors would like to acknowledge the outstanding efforts and invaluable contributions of the Jefferson Lab staff and all participating institutions that contributed to the success of this experiment. This work was supported in part by the the National Science Foundation (NSF) under grants PHY 1505616, 10011349, and 10014377, the U.S. Department of Energy (DOE), Office of Science, Office of Nuclear Physics under contract 89243126CSC000213. 
This work was furthermore supported in part by the University of South Carolina, the Chilean Comisi\'on Nacional de Investigaci\'on Cient\'ifica y Tecnol\'ogica (CONICYT), the Italian Istituto Nazionale di Fisica Nucleare, the French Centre National de la Recherche Scientifique, the French Commissariat \`{a} l'Energie Atomique, the Scottish Universities Physics Alliance (SUPA), the United Kingdom's Science and Technology Facilities Council, the National Research Foundation of Korea, the German Research Foundation (DFG), and the Skobeltsyn Nuclear Physics Institute and Physics Department at the Lomonosov Moscow State University. 

\end{acknowledgments}

\clearpage
\vfil

\bibliography{references}{}
\bibliographystyle{apsrev4-1}

\end{document}